\documentclass[10pt,times,a4paper,floatfix]{iopart}
\usepackage{epsfig}
\usepackage{graphics}
\usepackage{graphicx}
\usepackage{bm}
\usepackage{iopams}
\usepackage[dvipsnames]{xcolor}
\usepackage{bm}
\usepackage{times}
\usepackage{xspace}
\usepackage[varg]{txfonts}
\usepackage[normalem]{ulem} 
\usepackage[colorlinks]{hyperref}
\usepackage[caption=false]{subfig}
\usepackage{booktabs}

\definecolor{LinkColor}{rgb}{0.75, 0, 0}
\definecolor{CiteColor}{rgb}{0, 0.5, 0.5}
\definecolor{UrlColor}{rgb}{0, 0, 0.75}
\hypersetup{linkcolor=LinkColor}
\hypersetup{citecolor=CiteColor}
\hypersetup{urlcolor=UrlColor}
\usepackage{perpage}
\MakePerPage{footnote}

\newcommand{\dd}{\mathrm{d}}
\newcommand{\Pimr}{P_{\textsc{imr}}}
\newcommand{\Pin}{P_{\textsc{i}}}
\newcommand{\Prd}{P_{\textsc{mr}}}

\newcommand{\fisco}{f_\textsc{isco}}
\newcommand{\dMfbyMf}{\Delta M_f/\bar{M}_f}
\newcommand{\dafbyaf}{\Delta a_f/\bar{a}_f}

\newcommand{\ba}{\boldsymbol{a}}

\newcommand{\msun}{{M}_\odot}

\newcommand{\lalinferencenest}{\textsc{LALInferenceNest}\xspace}

\newcommand{\paperone}{Paper~I\xspace}

\begin{document}

\title[Testing general relativity using binary black holes]{Testing general relativity using gravitational wave signals from the inspiral, merger and ringdown of binary black holes}

\author{Abhirup~Ghosh$^1$, Nathan~K.~Johnson-McDaniel$^1$, Archisman~Ghosh$^{2,1}$, Chandra~Kant~Mishra$^{3,1}$, Parameswaran~Ajith$^{1,8}$, Walter~Del~Pozzo$^{4,5}$, Christopher~P.~L.~Berry$^5$, Alex~B.~Nielsen$^6$, Lionel~London$^7$}
\address{$^1$ International Centre for Theoretical Sciences, Tata Institute of Fundamental Research, Bangalore~560089, India}
\address{$^2$ Nikhef -- National Institute for Subatomic Physics, Science Park~105, 1098~XG Amsterdam, The~Netherlands}
\address{$^3$ Department of Physics, Indian Institute of Technology Madras, Chennai~600036, India}
\address{$^4$ Dipartimento di Fisica ``Enrico Fermi'', Universit\`a di Pisa, Pisa~I-56127, Italy}
\address{$^5$ School of Physics and Astronomy, University of Birmingham, Edgbaston, Birmingham, B15~2TT, United~Kingdom}
\address{$^6$ Max-Planck-Institut f\"ur Gravitationsphysik, Albert-Einstein-Institut, Callinstra{\ss}e~38, 30167 Hannover, Germany}
\address{$^7$ School of Physics and Astronomy, Cardiff University, The Parade, Cardiff, CF24~3AA, United~Kingdom}
\address{$^8$ Canadian Institute for Advanced Research, CIFAR Azrieli Global Scholar, MaRS Centre, West Tower, 661 University Ave., Suite 505, Toronto, ON M5G 1M1, Canada}


\begin{abstract}
Advanced LIGO's recent observations of gravitational waves (GWs) from merging binary black holes have opened up a unique laboratory to test general relativity (GR) in the highly relativistic regime. One of the tests used to establish the consistency of the first LIGO event with a binary black hole merger predicted by GR was the \emph{inspiral--merger--ringdown consistency test}. This involves inferring the mass and spin of the remnant black hole from the inspiral (low-frequency) part of the observed signal and checking for the consistency of the inferred parameters with the same estimated from the post-inspiral (high-frequency) part of the signal. Based on the observed rate of binary black hole mergers, we expect the advanced GW observatories to observe hundreds of binary black hole mergers every year when operating at their design sensitivities, most of them with modest signal to noise ratios (SNRs). Anticipating such observations, this paper shows how constraints from a large number of events with modest SNRs can be combined to produce strong constraints on deviations from GR. Using kludge modified GR waveforms, we demonstrate how this test could identify certain types of deviations from GR if such deviations are present in the signal waveforms. We also study the robustness of this test against reasonable variations of a variety of different analysis parameters. 
\end{abstract}

\section{Introduction}\label{sec:intro}
On September 14, 2015 the Advanced LIGO detectors~\cite{TheLIGOScientific:2014jea, TheLIGOScientific:2016agk} made the first observation of gravitational waves (GWs) from a binary black hole system~\cite{Abbott:2016blz}. The signal, termed GW150914, was inferred to be produced by the merger of two black holes (with masses $\sim 36 \msun$ and $29 \msun$) at a distance of $\sim 400~\mathrm{Mpc}$ from the Earth~\cite{TheLIGOScientific:2016wfe,TheLIGOScientific:2016pea}. This observation was followed by a second one on December 26, 2015~\cite{Abbott:2016nmj} and a third one on January 4, 2017~\cite{Abbott:2017vtc}. The second signal, termed GW151226, was produced by the coalescence of two black holes with masses $\sim 14 \msun$ and $8 \msun$, while the third signal GW170104 was produced by a binary black hole system more like GW150914's source (masses $\sim 31 \msun$ and $19 \msun$). These observations herald the beginning of a new branch of observational astronomy. Apart from providing the first direct evidence of the existence of GWs, these observations confirmed the existence of stellar mass black holes that are much more massive than previously thought ($\gtrsim 25 \msun$)~\cite{TheLIGOScientific:2016htt}. They also provided the first evidence of binary black holes that inspiral under GW emission and subsequently merge. These observations also enabled us to perform the first tests of general relativity (GR) in the highly relativistic and nonlinear regime of gravity~\cite{TheLIGOScientific:2016src, TheLIGOScientific:2016pea} --- a regime with binary orbital velocities $\sim 0.5\,c$, which is inaccessible by other astronomical observations and laboratory tests.

A handful of tests have been performed to check the consistency of the observed signals with those predicted by GR. In particular, due to the large masses of the black holes in GW150914, the observed signal consists of signatures from the inspiral and merger of the two black holes and the subsequent ringdown of the final black hole. This allowed us to perform several consistency tests making use of one or more phases of the coalescence: The first involved testing the consistency between the mass and spin of the final remnant, determined from the low-frequency (inspiral) and high-frequency (post-inspiral) parts of the observed signal~\cite{Ghosh:2016xx}. The second involved testing the consistency of the data after the peak of the observed signal (corresponding to the merger) with a quasi-normal mode (QNM) spectrum predicted by GR~\cite{1970Natur.227..936V,1971ApJ...170L.105P,Chandrasekhar441, Berti:2009kk}. The third involved bounding deviations from the GR predictions of the post-Newtonian (PN) coefficients describing the inspiral (and from phenomenological parameters describing the merger and ringdown) using parametrized waveform models~\cite{yunes2009fundamental,Mishra:2010tp,Agathos:2013upa}. The fourth involved constraining the amount of dispersion in the observed GW signal and converting it to a bound on the mass of the graviton~\cite{Will:1997bb}. In addition, the residuals after subtracting the best fit GR signal have been found to be consistent with detector noise.  Within the statistical uncertainties, these investigations provided no evidence for deviations from GR~\cite{TheLIGOScientific:2016src}. Some of these tests were repeated by using the second and third LIGO events, and combining the results from multiple events allowed us to improve the constraints on certain departures from GR predictions~\cite{TheLIGOScientific:2016pea,Abbott:2017vtc}. Of particular interest to this paper is the first test mentioned above that checks for the consistency of the properties of the remnant estimated from the inspiral and merger--ringdown parts of the observed signal, which we call the \emph{inspiral--merger--ringdown} (IMR) \emph{consistency test}~\cite{Ghosh:2016xx}. This paper provides further details of the formulation and implementation of this test that was used to constrain certain departures from GR using the LIGO events GW150914 and GW170104~\cite{Abbott:2017vtc}. 

The Advanced LIGO detectors, are not yet operating at than their design sensitivities~\cite{TheLIGOScientific:2014jea}, but are expected to approach design sensitivity over the next few years~\cite{Aasi:2013wya}. The Advanced Virgo~\cite{TheVirgo:2014hva} detector in Italy joined the Advanced LIGO detectors towards the end of the second observing run, and commissioning should bring it to design sensitivity on a similar timescale as the Advanced LIGO detectors. Several observing runs will be conducted over this period. Based on the rate of GW events so far~\cite{Abbott:2017vtc}, and the anticipated improvement in the sensitivity of the instruments, we expect that GW observations will become routine. The binary black hole event rate is expected to approach several hundred detections per year when the LIGO detectors operate at their design sensitivity~\cite{Abbott:2016nhf, TheLIGOScientific:2016pea}. The KAGRA detector~\cite{Aso:2013eba} in Japan is expected to join the international network of advanced GW detectors in the next few years, and the LIGO-India~\cite{LIGOIndiaProposal:2011} detector is planned to join the network around 2024~\cite{Aasi:2013wya}. These additional detectors will dramatically improve the sky coverage, source localization accuracy and the polarization extraction capability of the international GW network.

Upcoming GW observations will allow us to make significant improvements to the present constraints on deviations from GR, or potentially to detect such deviations --- either from a small number of particularly loud events or by combining constraints from a large number of moderate events. Thus, a careful characterization of the analysis pipelines and various systematic errors in the analysis is of vital importance, so that systematic errors in our analysis will not be mistaken for a true deviation from GR. In anticipation of the binary black hole events expected in the upcoming observational runs of Advanced LIGO and other ground-based observatories, we present here a detailed follow up of the analysis method of the IMR consistency test for binary black holes presented in our previous paper~\cite{Ghosh:2016xx}, which we will henceforth refer to as \paperone.

The IMR consistency test was developed based on the original idea by Hughes and Menou~\cite{Hughes:2004vw} proposed in the context of the space-based LISA observatory.\footnote{See also, Nakano, Tanaka and Nakamura~\cite{Nakano:2015uja} for a recent study in the context of a ground-based detector.} The key idea is to infer the mass and spin of the remnant black hole, using two different parts of the observed signal, and then to compare these independent estimates. We first estimate the initial masses and spins from the inspiral (low-frequency) part of the signal, which allows us to infer the mass and spin of the final black hole making use of fitting formulas calibrated to numerical relativity (NR) simulations of binary black holes. Next, we estimate the same parameters independently from the merger--ringdown (high-frequency) part of the signal and then compare the two estimates. If the signal is correctly described by the merger of a (quasicircular) binary of Kerr black holes in GR, which is implicit in the waveform models that we use to estimate the parameters and the fits that went in, one should expect the two estimates to be consistent with each other. On the other hand, if there is a departure from GR, depending on the exact nature of the departure, it can manifest as a discrepancy between the two estimates. In particular, if the energy and angular momentum radiated during the merger regime (where the gravity is extremely strong and nonlinear) differ significantly from the GR predictions for these quantities, one could expect a discrepancy between the two estimates in the final mass and spin. 

The original idea by Hughes and Menou~\cite{Hughes:2004vw} was to estimate the parameters purely from the early inspiral (well described by the PN approximation to GR) and the late ringdown (well described by a spectrum of QNMs). According to this, we would estimate the parameters of the binary from the early inspiral signal (where any departure from GR is presumably small) and use these estimates to predict the properties of the remnant black hole, which could be estimated independently from the QNM ringdown. However, such a test will be possible only using a small number of \emph{golden} events, where both the early inspiral and late ringdown are observed with high SNRs. While such a test might be possible in the future using LISA or third generation ground-based instruments, such tests are unlikely to be possible using the current generation of GW observatories. Meanwhile, Advanced LIGO is expecting to observe several hundred binary black hole events of moderate SNR in the coming years. Hence, tests of GR in the near future are going to progress through building evidence by combining multiple events. Our formulation of the test is geared in this direction. Indeed, the transition from inspiral regime to merger regime does not happen at a precise time or frequency. We use some reasonable choice of a cutoff frequency to separate the signal to the low-frequency (inspiral) and high-frequency (merger--ringdown) parts. As part of our studies, we show that the test is robust against variations of this cutoff frequency that still give large enough SNRs in for parameter estimation in both the low- and high-frequency portions independently.

Section~\ref{sec:method} contains a detailed description of the analysis method. We then show how a large number of GW signals from binary black holes can be combined to produce strong constraints on the deviations from GR (section~\ref{sec:GRsims}), and how a true deviation from GR might show up in our analysis (section~\ref{sec:modGRsims}). We also demonstrate the robustness of our analysis against reasonable choices of the frequency that is used to demarcate the inspiral and post-inspiral parts of the observed signal (section~\ref{sec:robustCutoff}), the choice of the waveform template (section~\ref{sec:robustWaveform}), fitting formulas for the mass and spin of the final black hole (section~\ref{sec:robustNRFit}), and the effects of precession and higher modes (section~\ref{sec:precHM}). We conclude and consider future work in section~\ref{sec:conclusions}.

\section{Testing the consistency between the inspiral, merger and ringdown}
\label{sec:method}
The GW signal from a spinning binary black hole with negligible orbital eccentricity is described by $15$ parameters (enumerated in, e.g., section~II~B of~\cite{Veitch:2014wba}). If the spins of the individual black holes are aligned (or antialigned) with the direction of the orbital angular momentum, then the number of parameters reduces to $11$. These parameters are typically estimated by means of Bayesian inference. Bayes' theorem states that the posterior probability distribution of a parameter set $\blambda := \{\lambda_i\}$ of a model hypothesis $\mathcal{H}$, given data $d$ and any other information $I$, is:
\begin{equation}
P(\blambda | d, \mathcal{H}, I) = \frac{P(\blambda | \mathcal{H}, I) \, \mathcal{L}(d | \blambda, \mathcal{H}, I)}{P(d|\mathcal{H}, I)},
\label{eq:Bayes_theorem}
\end{equation}
where $P(\blambda | \mathcal{H}, I)$ is the \emph{prior} probability of the parameter set $\blambda$ given $\mathcal{H}$ and $I$,  while $\mathcal{L}(d | \blambda, \mathcal{H}, I)$ is called the \emph{likelihood} function, which is the probability of observing the data $d$, given $\blambda$, $\mathcal{H}$ and $I$. What appears in the denominator is a normalization constant $P(d|\mathcal{H}, I) := \int P(\blambda | \mathcal{H}, I) \, \mathcal{L}(d | \blambda, \mathcal{H}, I) \, \dd \blambda$, called the marginal likelihood, or the \emph{evidence} of the hypothesis $\mathcal{H}$. 

If our hypothesis $\mathcal{H}$ is that the data contains a GW signal described by a GR waveform model $h(\blambda)$ and stationary Gaussian noise described by the power spectral density $S_n(f)$, then, as described in appendix~A of~\cite{Cutler:1994ys}, the likelihood function can be defined as:

\begin{equation}
\mathcal{L}(d | \blambda, \mathcal{H}, I) \propto \exp\big[-\frac{1}{2} \langle d - h(\blambda) \, | \, d -h(\blambda) \rangle \big],
\label{eq:likelihood}\end{equation}
where $\langle \cdot | \cdot \rangle$ is the following noise-weighted inner product
\begin{equation}
\langle B | C \rangle := 2\, \int_{f_\mathrm{low}} ^{f_\mathrm{cut}} \frac{\tilde{B}^*(f)\tilde{C}(f) + \tilde{B}(f)\tilde{C}^*(f)}{S_n(f)} \, \dd f.
\label{eq:nwip}
\end{equation}
Above, $\tilde{B}(f)$ denotes the Fourier transform of $B(t)$ and a $^*$ denotes complex conjugation. The limits of integration ${f_\mathrm{low}}$ and ${f_\mathrm{cut}}$ are dictated by the bandwidth of the detector sensitivity, the bandwidth of the signal, as well as the cutoff frequencies in our calculations (described below). Owing to the large dimensionality of the parameter set $\blambda$, the posterior distribution $P(\blambda | d, \mathcal{H}, I)$ in equation~(\ref{eq:Bayes_theorem}) is computed by stochastically sampling the parameter space using techniques such as Markov-Chain Monte Carlo~\cite{Gregory:2005} or nested sampling~\cite{skilling2006}. For this paper, we use the \lalinferencenest~\cite{Veitch:2009hd, Veitch:2014wba} code that provides an implementation of the nested sampling algorithm in the LSC Algorithms Library~\cite{lal} for computing the posterior distributions. 

Using the Bayesian framework described above, we start by computing the joint posterior probability distribution on the initial masses and dimensionless spins $P(m_1, m_2, \ba_1, \ba_2\,|\,d)$\footnote{Here onwards, we drop $\mathcal{H}, I$ from the posteriors to simplifying the notation. Also, unless otherwise noted, the masses we consider are the ones in the detector frame, including the cosmological redshift~\cite{Krolak:1987ofj}.} of the binary black hole system (with $m_1 \geq m_2$ by convention). This is done by marginalizing the posterior over the remaining parameters describing the signal. Assuming quasicircular inspirals, fitting formulas calibrated to NR simulations then give us predictions of the mass $M_f$ and dimensionless spin magnitude $a_f$ of the final (remnant) black hole as functions of the initial masses and spins,
\begin{equation}
M_f = M_f(m_1, m_2, \ba_1, \ba_2)\,, \hspace{1 cm} a_f = a_f(m_1, m_2, \ba_1, \ba_2),
\label{eq:finalmassspin}
\end{equation}
which allow us to obtain the posterior probability distribution $P(M_f,  a_f\,|\,d)$ of the final mass and spin. 

\begin{figure*}
\centering 
\includegraphics[width=0.80\textwidth]{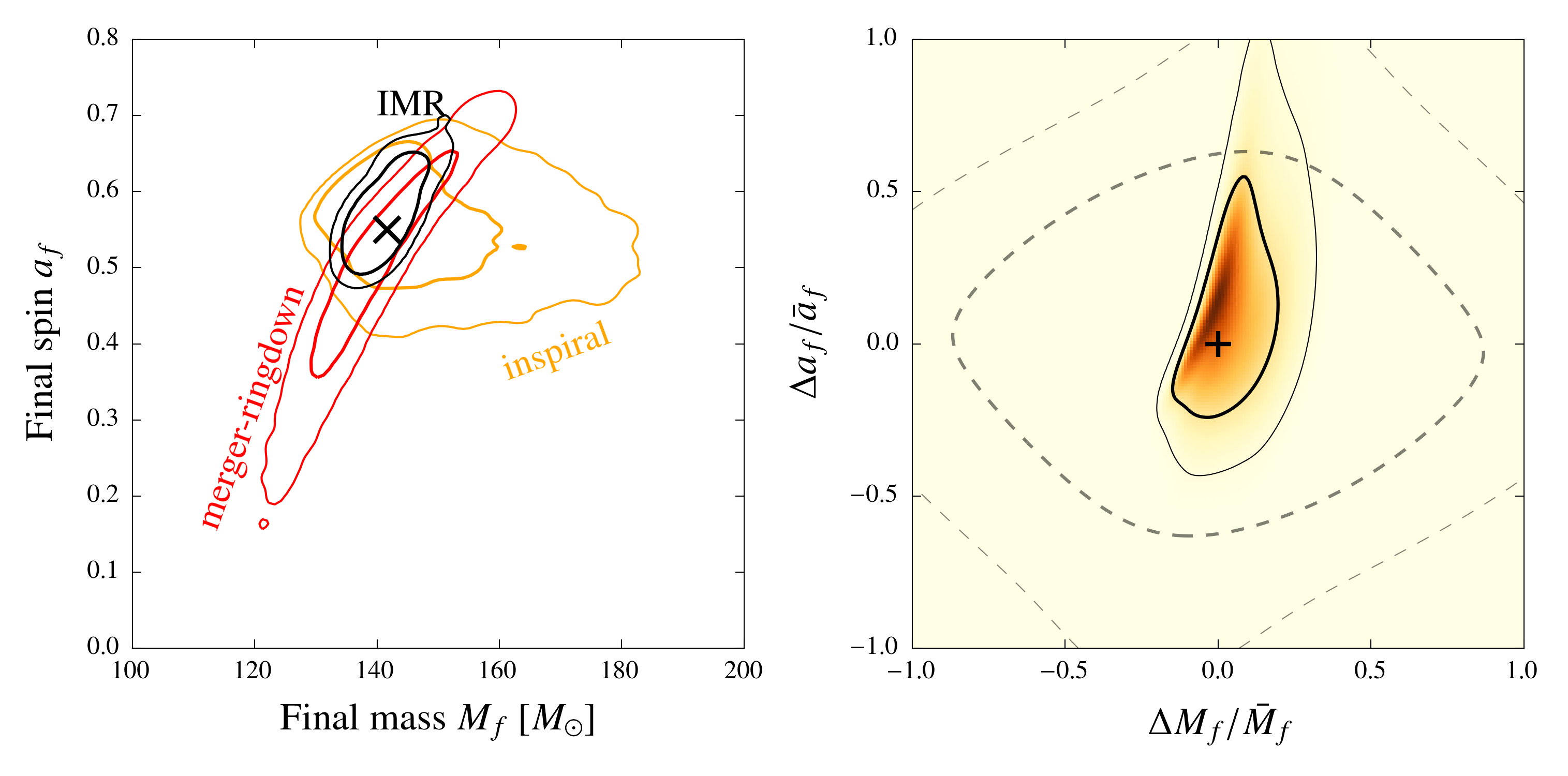}
\caption{The left panel shows the $68\%$ and $95\%$ credible regions of the posterior distributions $\Pin(M_f,a_f)$ and $\Prd(M_f,a_f)$ of the (redshifted) mass and spin of the final black hole estimated from the inspiral and merger--ringdown parts of a simulated GR signal, respectively. Also shown is the posterior $\Pimr(M_f, a_f)$ estimated from the full IMR signal. The simulated GR signal (using  \textsc{SEOBNRv2\_ROM\_DoubleSpin}~\cite{Purrer:2015tud,Taracchini:2013rva}) is from a binary black hole with $(m_1, m_2) = (60.7, 86.7) M_{\odot}$, and aligned spins $(a_1, a_2) = (0.65, -0.77)$ at a distance of $860~\mathrm{Mpc}$, observed by the $3$-detector Advanced LIGO--Virgo network. The corresponding value of the final mass and spin is indicated by a black cross. The right panel shows the posterior $P(\epsilon, \sigma)$ on the parameters $\epsilon := \dMfbyMf$ and  $\sigma := \dafbyaf$ that describe the deviation from GR for the same case as the left panel. Dashed lines show the 68\% and 95\% isoprobability contours of the prior used to compute the posterior, which corresponds to a uniform prior in component masses. The posterior is consistent with the GR value, which is marked by a `+' sign.} 
\label{fig:single_posterior}
\end{figure*}

First, we estimate the posterior $\Pimr(M_f,  a_f\,|\,d)$ using the full observed signal. We choose to demarcate the inspiral and post-inspiral parts of the signal using the $m = 2$ mode GW frequency $\fisco$ of the innermost stable circular orbit (ISCO) of the remnant Kerr black hole~\cite{Bardeen:1972fi}, with mass and spin given by the median values of $\Pimr(M_f,  a_f\,|\,d)$. We estimate the same parameters from the data containing only the inspiral (low-frequency) part of the observed signal; that is, by setting $f_\mathrm{cut} = \fisco$ in equations~(\ref{eq:likelihood}) and (\ref{eq:nwip}). This allows us to compute the posterior distribution $\Pin(M_f, a_f\,|\,d)$ of the mass and spin of the remnant purely from the inspiral part of the signal. Similarly, using only the merger--ringdown (high-frequency) part of the observed signal [by setting $f_\mathrm{low} = \fisco$ in equations~(\ref{eq:likelihood}) and  (\ref{eq:nwip})], we can get yet another estimate of the posterior $\Prd(M_f, a_f\,|\,d)$ of the mass and spin of the remnant. If the observed signal is well described by GR, the two independent estimates $\Pin(M_f, a_f\,|\,d)$ and $\Prd(M_f, a_f\,|\,d)$ have to be consistent with each other as well as with the estimate $\Pimr(M_f, a_f\,|\,d)$ using the full data (see, e.g., the left plot in figure~\ref{fig:single_posterior}).

To quantify the consistency of the observed signal with a binary black hole system predicted by GR, we define two parameters that describe the fractional difference between the two estimates of the remnant's mass and spin 
\begin{equation}
\epsilon := \frac{\Delta M_f}{\bar{M}_f}, ~~~~~ \sigma:= \frac{\Delta a_f}{\bar{a}_f},
\label{eq:epsilon_sigma}
\end{equation}
where 
\begin{equation}
\Delta M_f := M^\textsc{i}_f - M^\textsc{mr}_f, ~~~~~~ \Delta a_f := a^\textsc{i}_f - a^\textsc{mr}_f,
\end{equation}
and
\begin{equation}
\bar{M}_f := \frac{M^\textsc{i}_f + M^\textsc{mr}_f}{2}, ~~~~~~ \bar{a}_f := \frac{a^\textsc{i}_f + a^\textsc{mr}_f}{2}.
\label{eq:averages}
\end{equation}
In the absence of departures from GR, we expect the posterior $P(\epsilon, \sigma\,|\,d)$ of $\{\epsilon, \sigma\}$ to be consistent with zero (within the expected statistical fluctuations due to noise).\footnote{We used the full IMR posteriors for $M_f$ and $a_f$ to normalize the fractional parameters in \paperone. We have switched to using the average of the inspiral and merger--ringdown posteriors here so that we only are using posteriors from the two independent measurements, and not mixing in ones from the case where the parameters describing the inspiral and merger--ringdown parts of the waveform are not allowed to vary independently.} \ref{app:posterior_derivation} describes the calculation of $P(\epsilon, \sigma\,|\,d)$ from the posteriors $\Pin(M_f,a_f\,|\,d)$ and $\Prd(M_f,a_f\,|\,d)$. An example of the posterior distribution $P(\epsilon,\sigma\,|\,d)$ from a simulated GR signal is shown in figure~\ref{fig:single_posterior} (right plot).

Finally, if we assume that $\{\epsilon, \sigma\}$ take the same values in multiple events, the posterior from one event could be treated as the prior for computing the posterior for the second event (see section~\ref{sec:caveats} for a discussion on the validity of this assumption). Such a combined posterior can more sensitively identify certain deviations from GR. In practice, the posteriors $P_j(\epsilon, \sigma\,|\,d_j)$ from $N$ such observations can be combined to produce an overall posterior in the following way: 
\begin{equation}
P(\epsilon, \sigma\,|\,\{d_j\}) = P(\epsilon, \sigma) ~ \prod_{j=1}^{N} \frac{P_j(\epsilon, \sigma\,|\,d_j)}{P_j(\epsilon, \sigma)},
\end{equation}
where $P(\epsilon,\sigma)$ is the overall prior distribution on $\epsilon$ and $\sigma$ and $P_j(\epsilon, \sigma)$ is the prior distribution used to compute the posterior $P_j(\epsilon, \sigma\,|\,d_j)$. 

\subsection{Choice of priors}

Given our poor understanding of the astrophysical distribution of source parameters (such as the component masses and spins), we use uninformative priors for the parameter estimation~\cite{Veitch:2014wba,TheLIGOScientific:2016wfe}. Sources are assumed to be uniformly distributed in volume and isotropically oriented. We use uniform priors in component masses $m_{1,2} \in [1,300] \msun$, where $m_1 \geq m_2$, and a uniform prior in the spin magnitudes $|\ba_{1,2}| \in [0, 0.98]$. We use isotropic priors on the spin orientation when precessing templates are used in the parameter estimation. When nonprecessing templates are used, the priors are set in such a way that the dimensionless spin projection $\ba_{1,2} \cdot \hat{\mathbf{L}}$ onto the orbital angular momentum $\mathbf{L}$ has a uniform distribution in $[-0.98, 0.98]$. These choices induce a non-uniform prior in $\{\epsilon,\sigma\}$, which is shown by the dashed contours in Figure~\ref{fig:single_posterior} (right panel). 

\section{Simulations of GR signals in Gaussian noise}
\label{sec:GRsims}

\begin{figure}
\centering 
\includegraphics[width=2.5in]{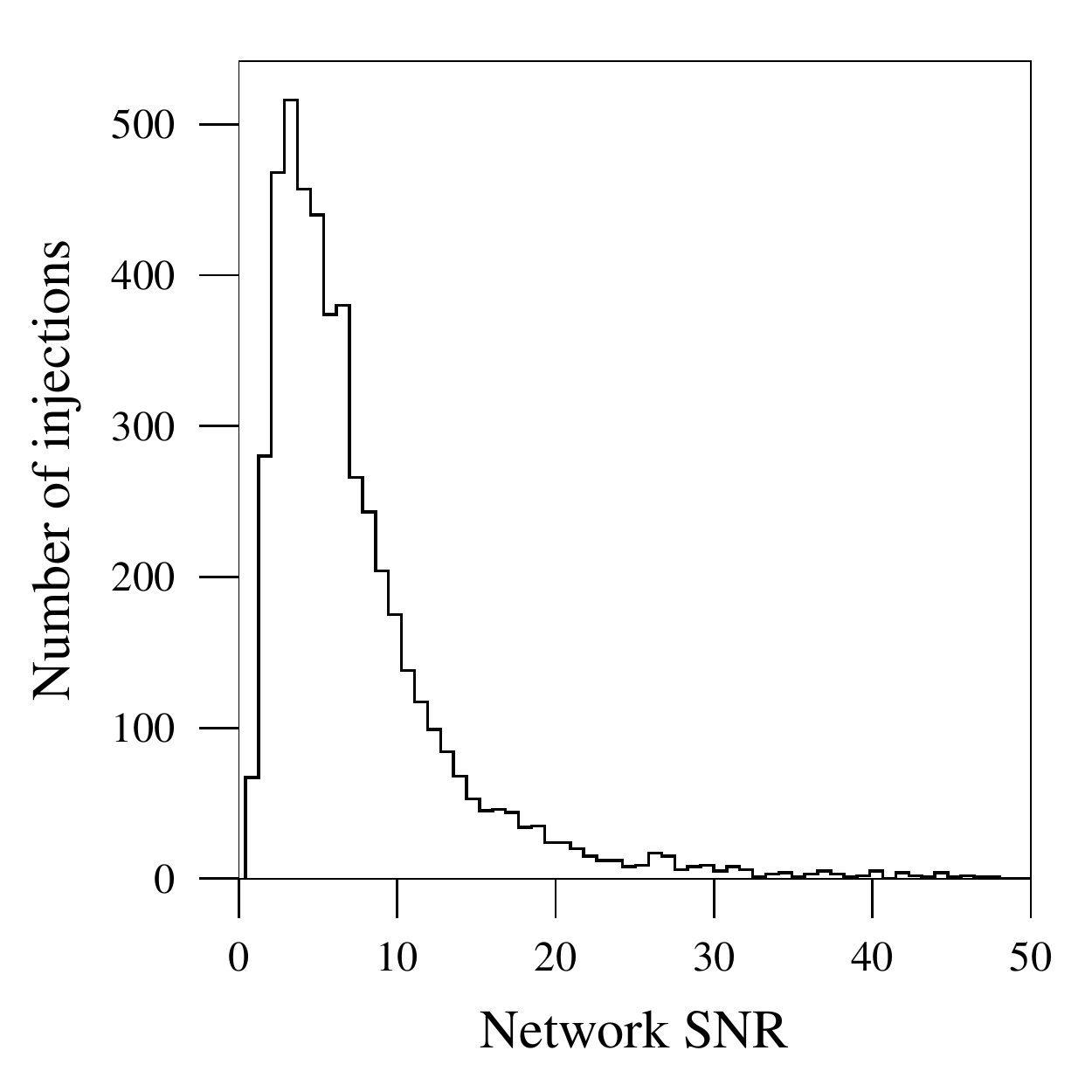}
\caption{Distribution of SNR of Gaussian noise simulations of GW signals from binary black hole coalescences uniformly distributed in a comoving four-volume between redshifts $[0,2]$, initial source-frame component masses and component spin magnitudes uniformly distributed between $[10, 80] M_{\odot}$, and $[0, 0.98]$.} 
\label{fig:snr_distribution}
\end{figure}

First, we estimate the expected constraints that we can place on parameters describing deviations from GR, when the actual signals are well described by GR. We inject simulated GW signals modeling inspiral, merger and ringdown of binary black holes (based on GR, modeled by the \textsc{SEOBNRv2\_ROM\_DoubleSpin} approximant~\cite{Purrer:2015tud,Taracchini:2013rva}) into colored Gaussian noise with the design power spectrum of the Advanced LIGO detectors in the high-power, zero-detuning configuration~\cite{AdvLIGOPSD}. Although we perform the analysis assuming the $3$-detector Advanced LIGO--Virgo network, for simplicity, we assume that Advanced Virgo has the same noise power spectrum as the Advanced LIGO detectors. All detectors are assumed to have a low frequency cutoff $f_\mathrm{low} = 10~\mathrm{Hz}$. We simulate a population of binary black holes that is uniformly distributed in comoving four-volume in the redshift interval $z = [0,2]$, with isotropic orientations (see Section~II~C of~\cite{Ghosh:2015jra} for more details). Binaries have (source-frame) component masses uniformly distributed in the range $m^\mathrm{s}_{1,2} = [10, 80] M_{\odot}$ and component spin magnitudes uniformly distributed in the range $[0, 0.98]$. The GW signals are redshifted so that the effective masses describing the signal in the detector frame become $m_{1,2} = m^\mathrm{s}_{1,2} (1+z)$~\cite{Krolak:1987ofj}. Black hole spins are always assumed to be aligned/antialigned with the orbital angular momentum, as is required by the waveform model used: This is not an overly restrictive assumption, because the spin components along the orbital angular momentum have the dominant effect on the binary's dynamics and GW emission~\cite{Schmidt:2012rh}. Figure~\ref{fig:snr_distribution} shows the distribution of the network SNR of the injected population of binaries. For our analysis, we consider signals with network SNR greater than $8$ for both the inspiral and merger--ringdown parts, using the Kerr ISCO frequency to split the signal.\footnote{Here, for simplicity, we use the Kerr ISCO frequency computed from the injected values of masses and spins to split the signal. In an actual observation, we use the median estimate of the Kerr ISCO frequency estimated from the observed IMR signal.} In addition, we only consider binaries with total redshifted mass less than $150 M_\odot$, so that the observed signals contain imprints of the inspiral, merger and ringdown. Hence, starting from an initial population of $5000$ injections, the final analysis is performed on around $150$ events which survive the above thresholds on SNR and total redshifted mass.

On the injections that pass the SNR and mass thresholds, we perform the analysis described in section~\ref{sec:method}, using \textsc{SEOBNRv2\_ROM\_DoubleSpin} as our template for parameter estimation and the final mass and spin fits from Healy~\emph{et al.}~\cite{Healy:2014yta}. Parameter estimation using \lalinferencenest\ is performed using priors that are uniform in component masses in the interval $m_{1,2} \in [1, 300] M_\odot$ and component (aligned/antialigned) spins in the interval $[0, 0.98]$. This corresponds to a non-uniform prior in the deviation parameters $\{\epsilon, \sigma\}$ that is centered around the GR value of $\{0, 0\}$. See figure~\ref{fig:single_posterior} for an example result. The left panel shows the estimated posterior distributions of the final mass and spin from the inspiral and post-inspiral parts of the data, along with that estimated from the full data. The right panel shows the posterior distribution on the parameters describing deviations from GR, along with the prior used for computing this posterior. 

\begin{figure}[tbh]
\centering
\includegraphics[height=2.1in]{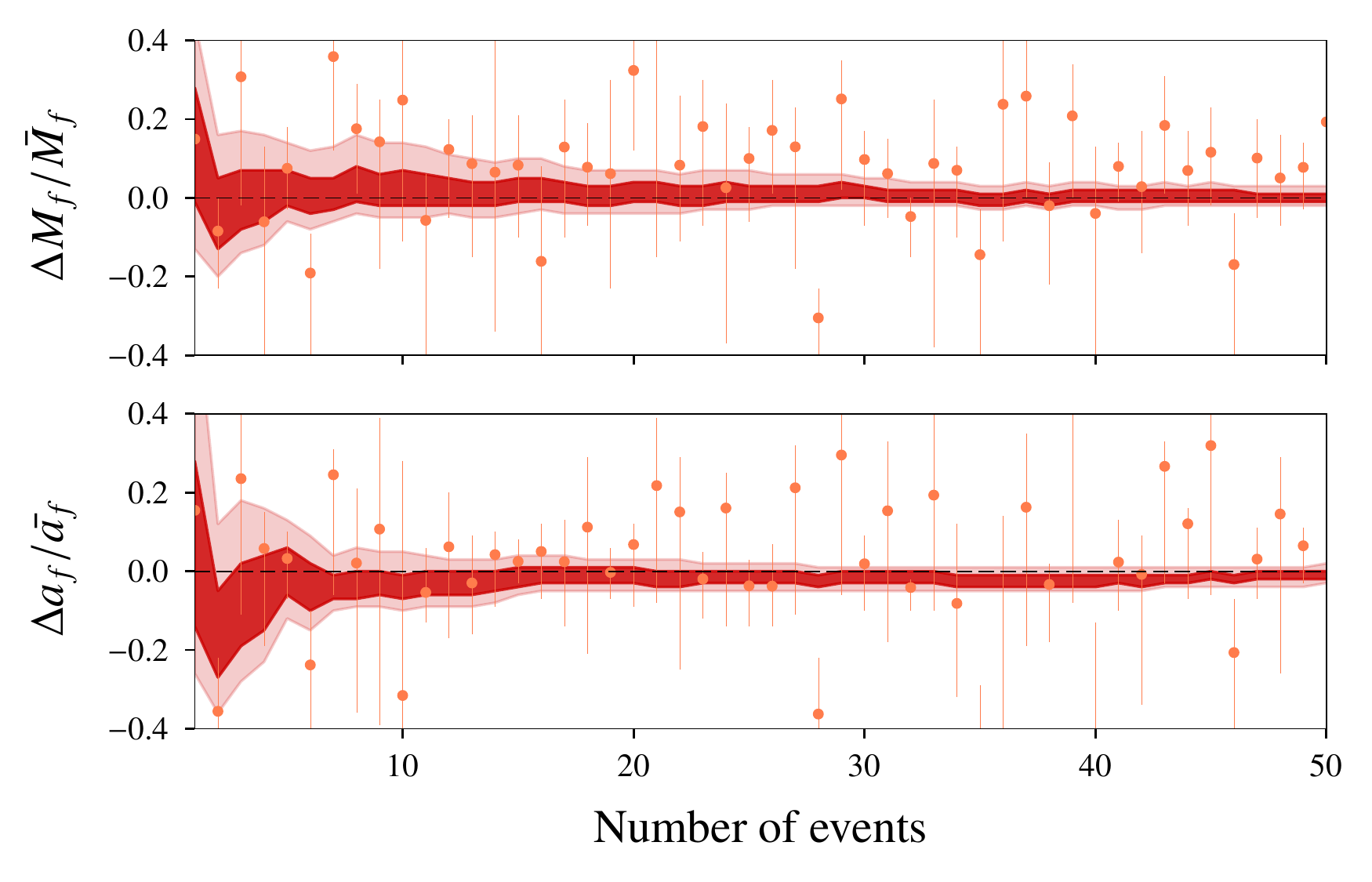}
\includegraphics[height=2.1in]{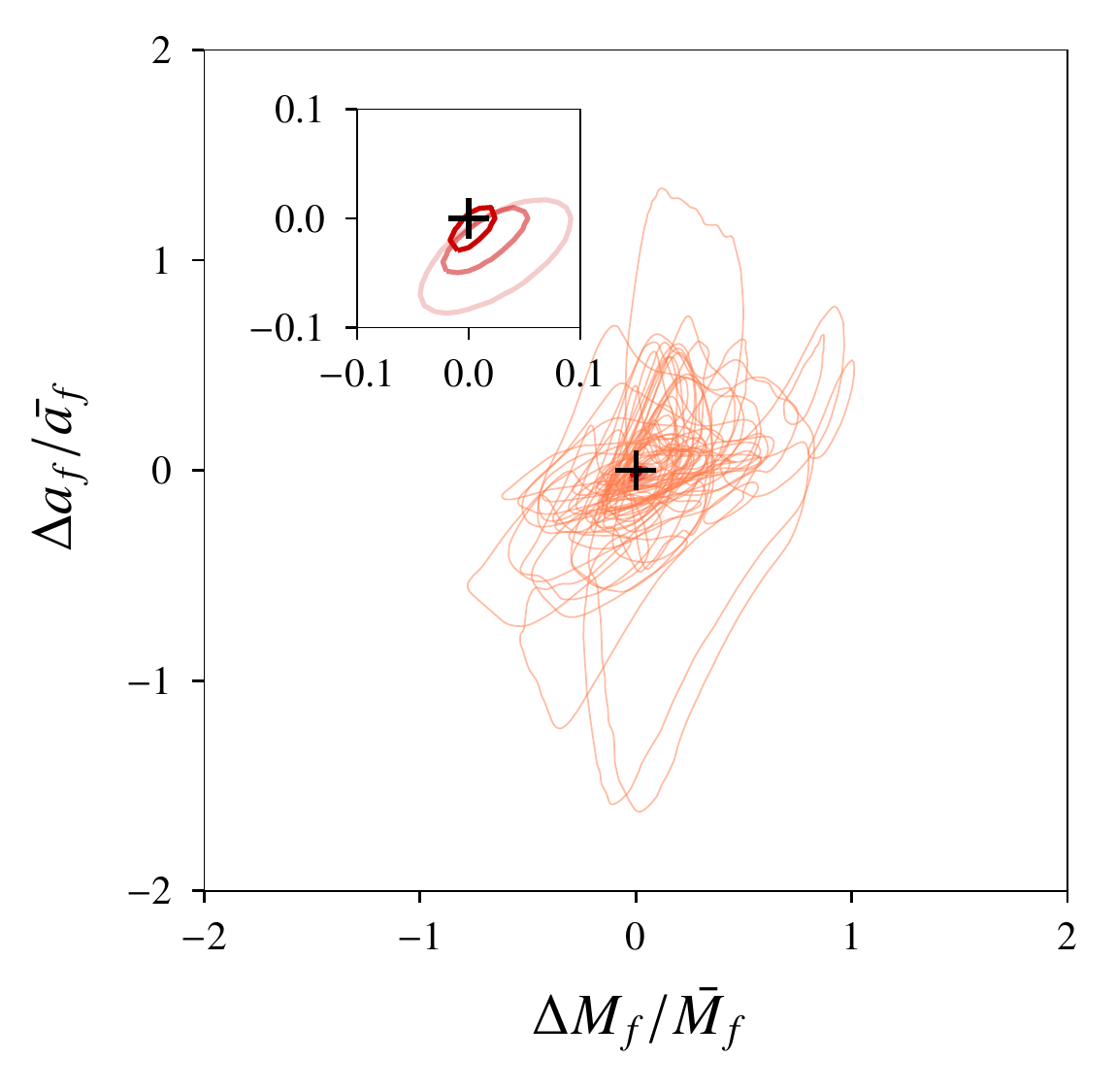}
\includegraphics[height=2.1in]{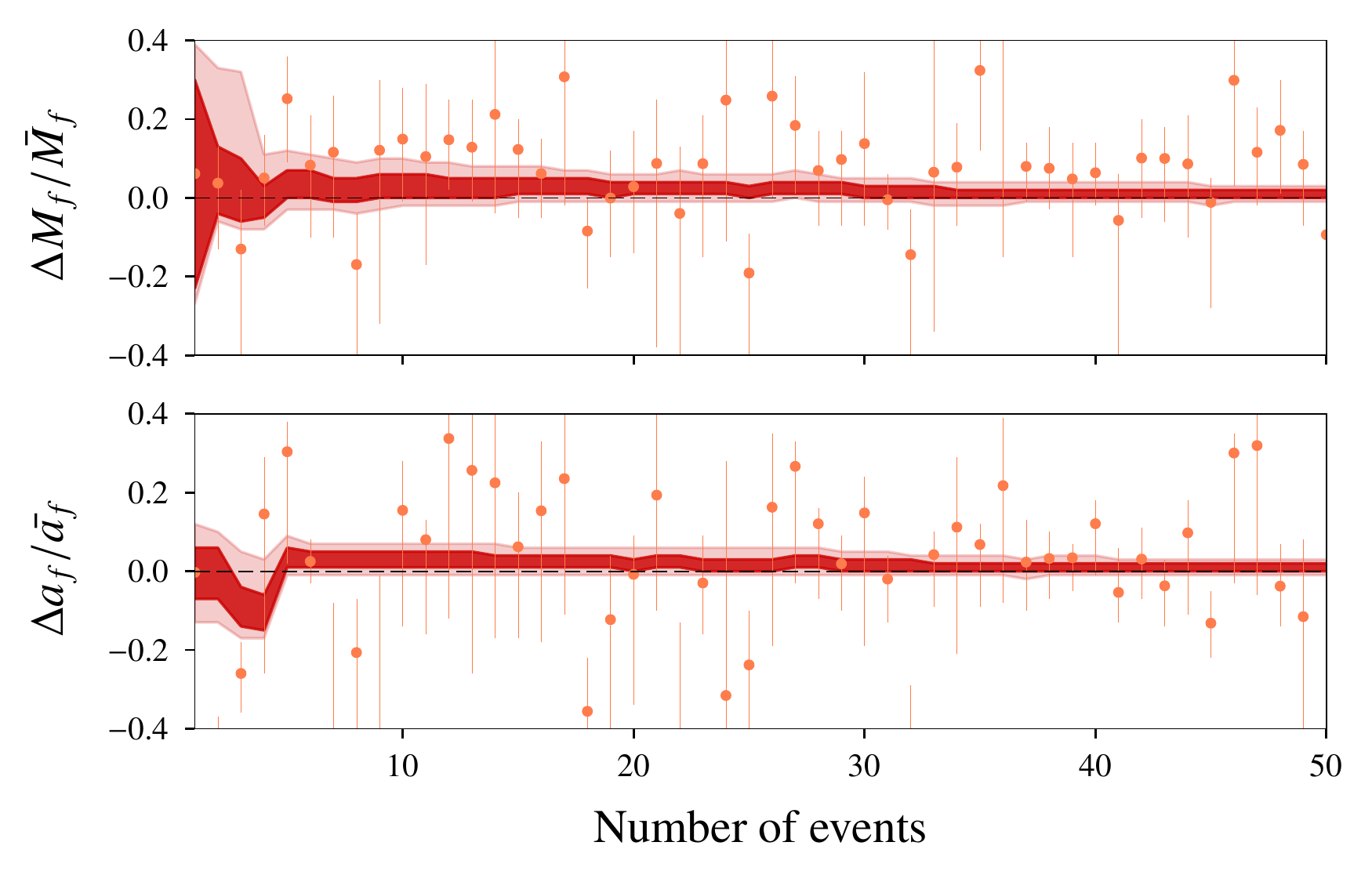}
\includegraphics[height=2.1in]{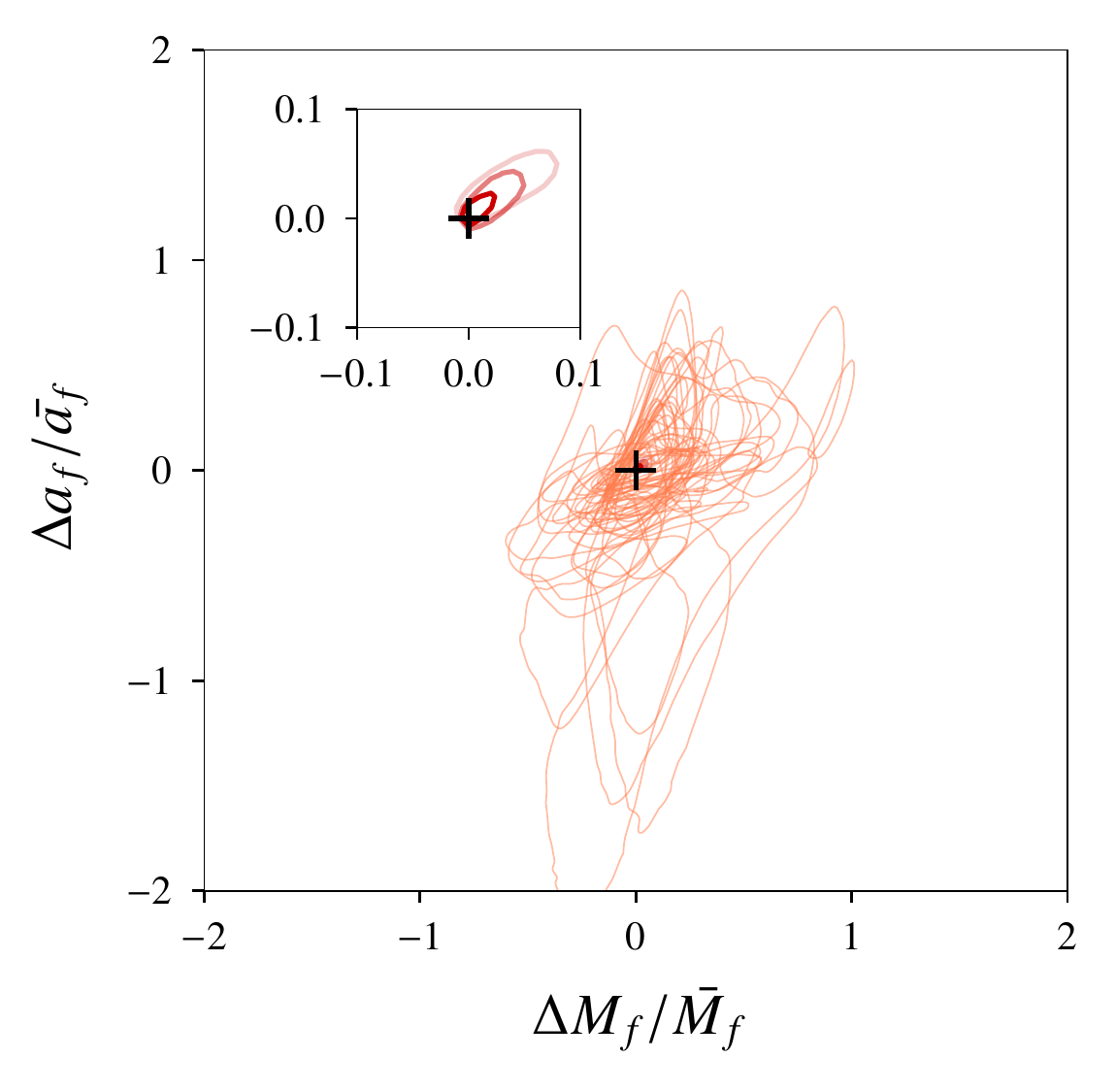}
\includegraphics[height=2.1in]{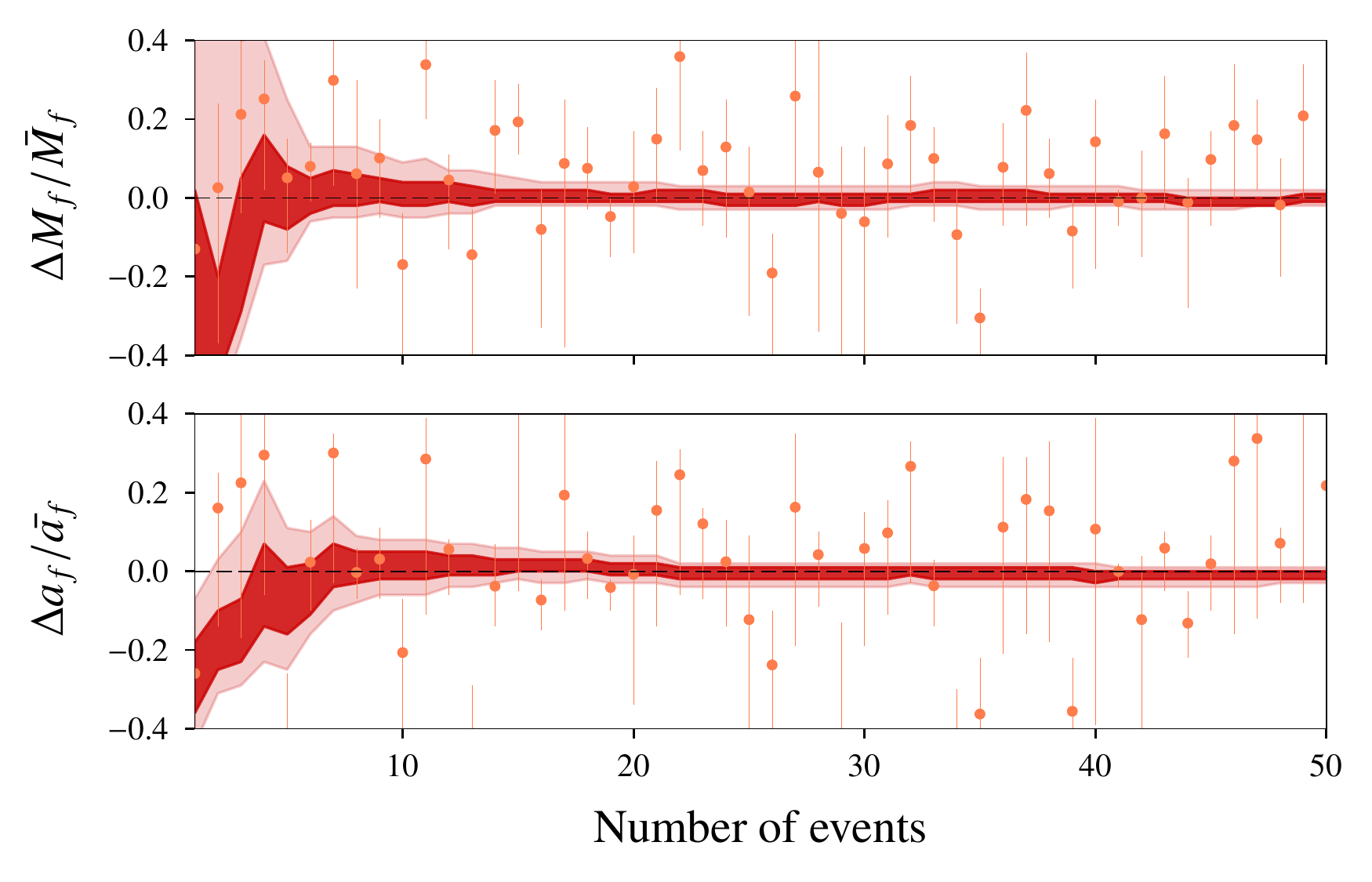}
\includegraphics[height=2.1in]{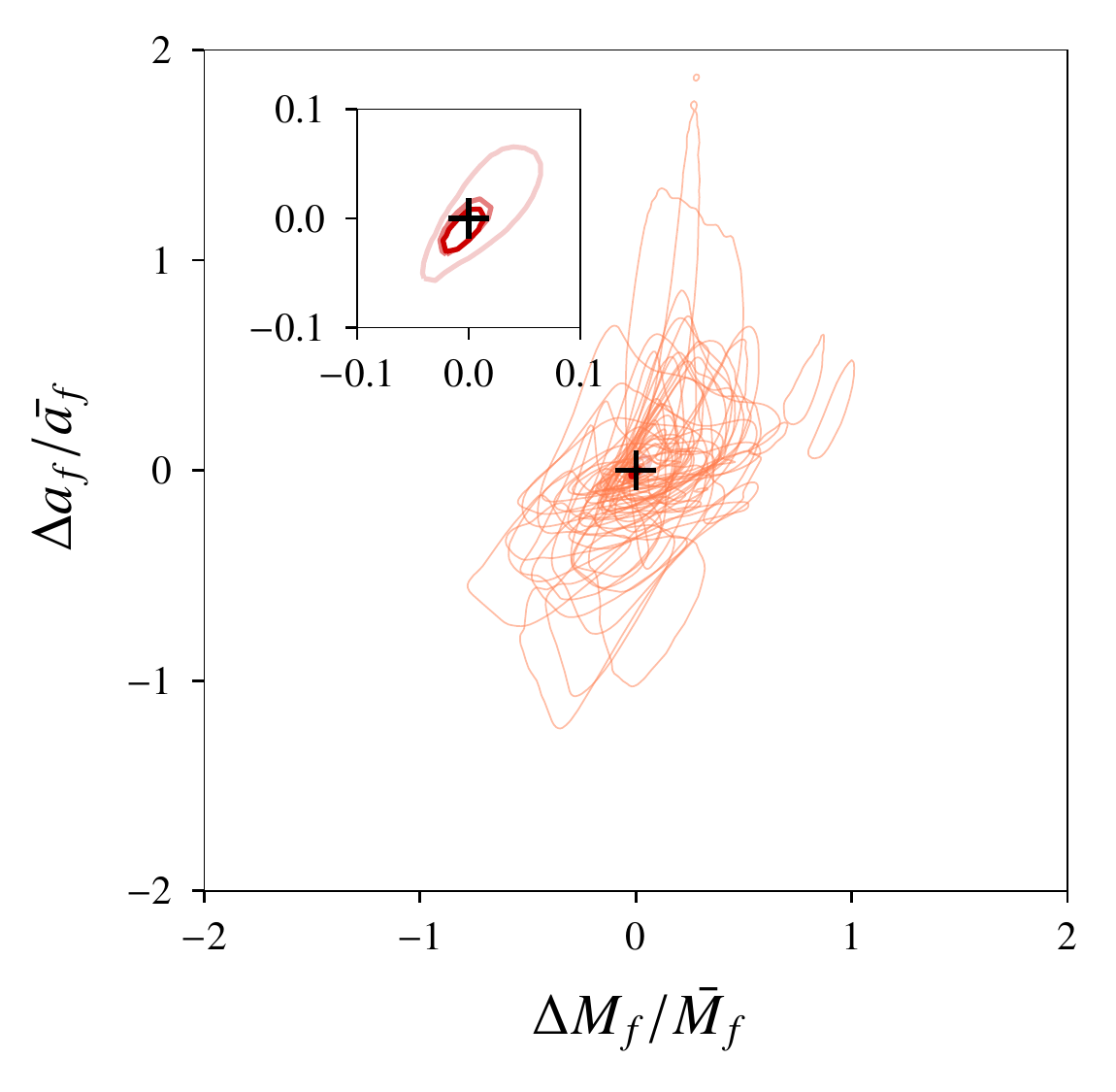}
\caption{\emph{Left:} In each plot shaded regions show the $68\%$ and $95\%$ credible intervals on the combined posteriors on $\epsilon := \dMfbyMf$, $\sigma := \dafbyaf$ from multiple observations of GR signals, plotted against the number of binary black hole observations. The GR value ($\epsilon = \sigma = 0$) is indicated by horizontal dashed lines. The mean value of the posterior from each event is shown as an orange dot along with the corresponding $68\%$ credible interval as an orange vertical line. Posteriors on $\epsilon$ are marginalized over $\sigma$, and vice versa. \emph{Right}:  The thin orange contours show the $68\%$ credible regions of the individual posteriors on the $\epsilon, \sigma$ computed from the same events. The GR value ($\epsilon = \sigma = 0$) is indicated by the black + sign. \emph{Right inset:} The red contours show the $68\%$ credible regions on the combined posterior from 5, 10 and 25 events (with increasing shades of darkness). The GR value ($\epsilon = \sigma = 0$) is indicated by the black + sign. Different rows correspond to different catalogs of 50 randomly chosen events from a total of $\sim 100$ simulated events.}
\label{fig:jointpost_gr}
\end{figure}

Constraints on the deviations expected from single observations of binary black holes by second-generation ground-based detectors are expected to be modest, due to the large statistical errors from the relatively small SNRs expected for most systems (see, e.g., figure~\ref{fig:snr_distribution}). However, the constraints can be improved by combining the likelihoods of $\{\epsilon , \sigma\}$ from multiple events.\footnote{We expect the combined errors to go down roughly as $N^{-1/2}$, where $N$ is the number of events while the individual errors to go down as $\rho^{-1}$, where $\rho$ is the SNR of the event. Thus, combined constraints from $N$ events with roughly same SNR should be comparable to the constraints from a single event with an SNR of $\rho \, N^{1/2}$.} Figure~\ref{fig:jointpost_gr} shows the combined posteriors $P(\epsilon , \sigma)$ as a function of the number of simulated events. The constraints on the deviation parameters $\{\epsilon , \sigma\}$ become narrower when multiple events are combined. These constraints are within the reach of Advanced LIGO--Virgo observations in the next few years~\cite{Abbott:2016nhf, TheLIGOScientific:2016pea}. 

From the posterior distribution $P(\epsilon, \sigma)$, one can obtain the credible level corresponding to the GR value of the deviation parameters, i.e., $\{\epsilon,\sigma\}=\{0,0\}$. The credible level for a certain set of parameters $\{\epsilon_0, \sigma_0\}$ is the total probability enclosed within an isoprobability contour passing through $\{\epsilon_0, \sigma_0\}$:
\begin{equation}
\label{eq:credible}
{\rm credible\ level} = \int_{\{\epsilon,\sigma\} {\rm \ where\ } {P(\epsilon,\sigma)>P(\epsilon_0,\sigma_0)}}\,P(\epsilon,\sigma)\,\dd\epsilon\,\dd\sigma.
\end{equation}
See, e.g., equation~(42) in~\cite{Veitch:2014wba} for a general definition. The area enclosed within the isoprobability contour is the credible area (or credible interval in $1$ dimension). A lower credible level for the GR parameters indicates a greater agreement with GR. However, this GR credible level is not expected to be zero even when the data is correctly described by GR, because the peak of posterior on the deviation parameters $\{\epsilon,\sigma\}$ is randomly shifted away from $\{0,0\}$ in the presence of noise. 

\begin{figure*}[tbh]
\centering
\includegraphics[width=\textwidth]{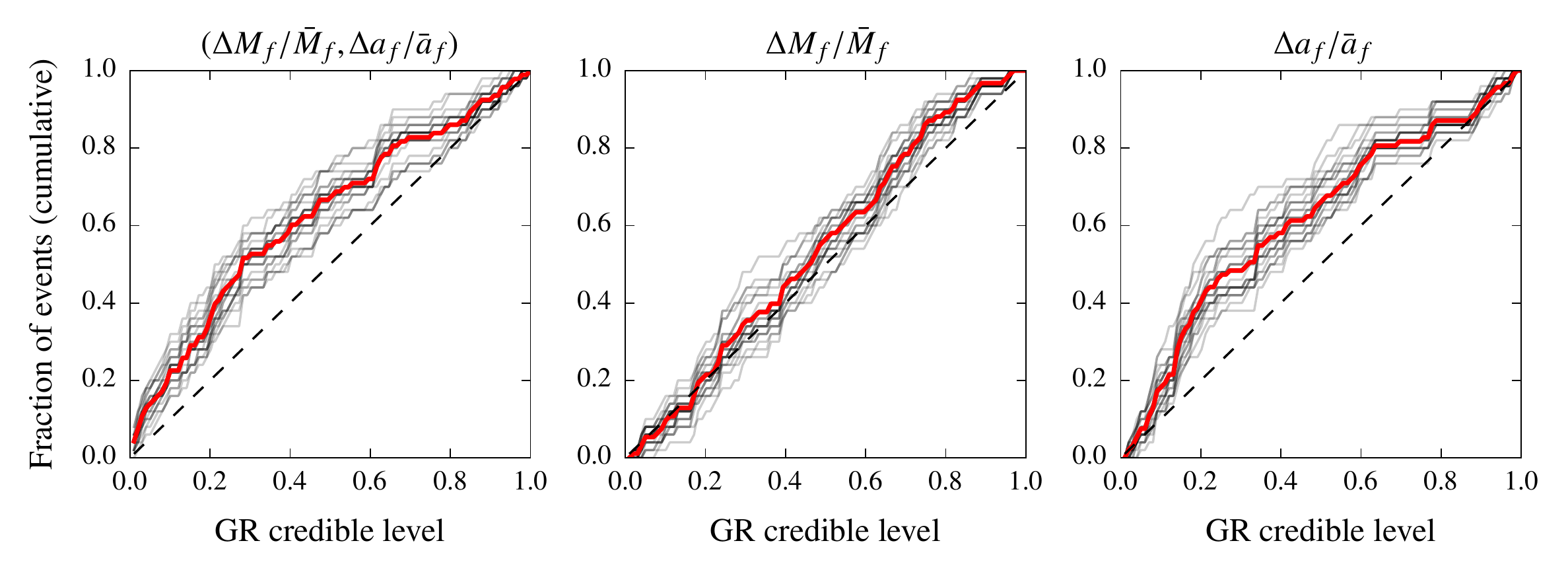}
\caption{A $p$ vs $p$ plot for the deviation parameters ($\epsilon := \Delta M_f/M_f$, $\sigma := \Delta a_f/a_f$)  computed from the set of GR injections described in section~\ref{sec:GRsims}. In each plot, the horizontal axis indicates a credible level, and the vertical axis indicates the fraction of events with the deviation parameters below the given credible level. In each plot, the gray lines are computed for $50$ different subsets of the full population of $\sim 100$ events, and the red line correspond to the same computed using all the $\sim 100$ events. The left plot corresponds to credible intervals computed from the $2$-dimensional posteriors $P(\epsilon, \sigma|d)$, while the middle and right plots correspond to marginalized posteriors on $\epsilon$ and $\sigma$, respectively. The diagonal lines indicate the case where the Bayesian credible levels match the frequentist confidence levels.}
\label{fig:GR-ppplots}
\end{figure*}

If the parameter values $\{\epsilon,\sigma\}$ were sampled from the prior distribution used to compute their posterior, we would expect the $p$ credible interval of the posterior to include the true value approximately $p$ of the time, for the case of an appropriately chosen prior~\cite{datta2005probability,Cook:2006}. However, for the GR injections that we performed, we do not expect the true value of $\{\epsilon = 0, \sigma = 0\}$ to be found at credible level $p$ for a fraction $p$ of events. This is because of our prior distribution for $\{\epsilon,\sigma\}$ does not match the intrinsic distribution (which would be a delta function): we are allowing for the possibility of mismatches between inspiral and merger--ringdown even though this is not the case for the injected GR signals. To demonstrate this for our set of GR injections, in figure~\ref{fig:GR-ppplots} we show the fraction of events for which the (true) GR point is found within the $p$ credible region. From figure~\ref{fig:GR-ppplots}, we see the true GR value is often near the peak of the distribution, meaning that it is found within the $p$ credible region for more than a fraction $p$ of events.\footnote{For the marginalised $\epsilon$ distribution, we see that the true GR value is approximately found within the $p$ credible interval a fraction $p$ of the time. In this case, the scatter due to noise is sufficient to balance the mismatch in the prior used.} In order to make a frequentist statement about the significance of the measurement of non-zero $\epsilon$ or $\sigma$ (how common such a deviation is if GR were correct), a similar study would need to be done for the region of mass and spin space of interest in order to calibrate expectations.

\setcounter{footnote}{0}
\section{Simulations of modified GR signals in Gaussian noise}
\label{sec:modGRsims}

\subsection{Generation of modified GR waveforms}
\label{sec:mod_gr_waveforms}

We also test the sensitivity of our pipeline towards certain kinds of deviations from GR, using signals that differ slightly from the predictions of GR. Specifically, we generate kludge modified GR waveforms that are qualitatively similar to binary black hole waveforms in GR, but differ in their energy and angular momentum loss as a function of frequency. These waveforms are similar enough to GR binary black hole waveforms that they would likely be detected with a standard detection pipeline, as we discuss below. However, we demonstrate that the consistency test described in this paper is able to identify such deviations from GR by combining the posteriors from multiple observations.  

We start from the nonspinning effective one-body (EOB) waveform model given in~\cite{Damour:2012ky}, which is available as Matlab code at IHES~\cite{IHES_EOB_url}, and modify the GW flux to yield our kludge modified GR waveform. Specifically, we multiply all the modes of the waveform that first enter the flux at second PN order [i.e.\ $O(v^4/c^4)$, where $v$ is the binary's orbital velocity], viz the $(\ell,m) = (3,\pm 2)$, $(4,\pm 4)$, and $(4,\pm 2)$ modes, by a constant factor $\sqrt{\alpha_2}$, so that their contributions to the flux are multiplied by $\alpha_2$. This also modifies the flux at $3$PN and all higher PN orders. We start with the modes that first enter at $2$PN rather than the ones that first enter at $1$PN since the latter modes vanish for equal-mass nonspinning binaries, and we still want a nonvanishing modification in this limit.

\begin{figure*}[tbh]
\centering
\includegraphics[width=0.7\textwidth]{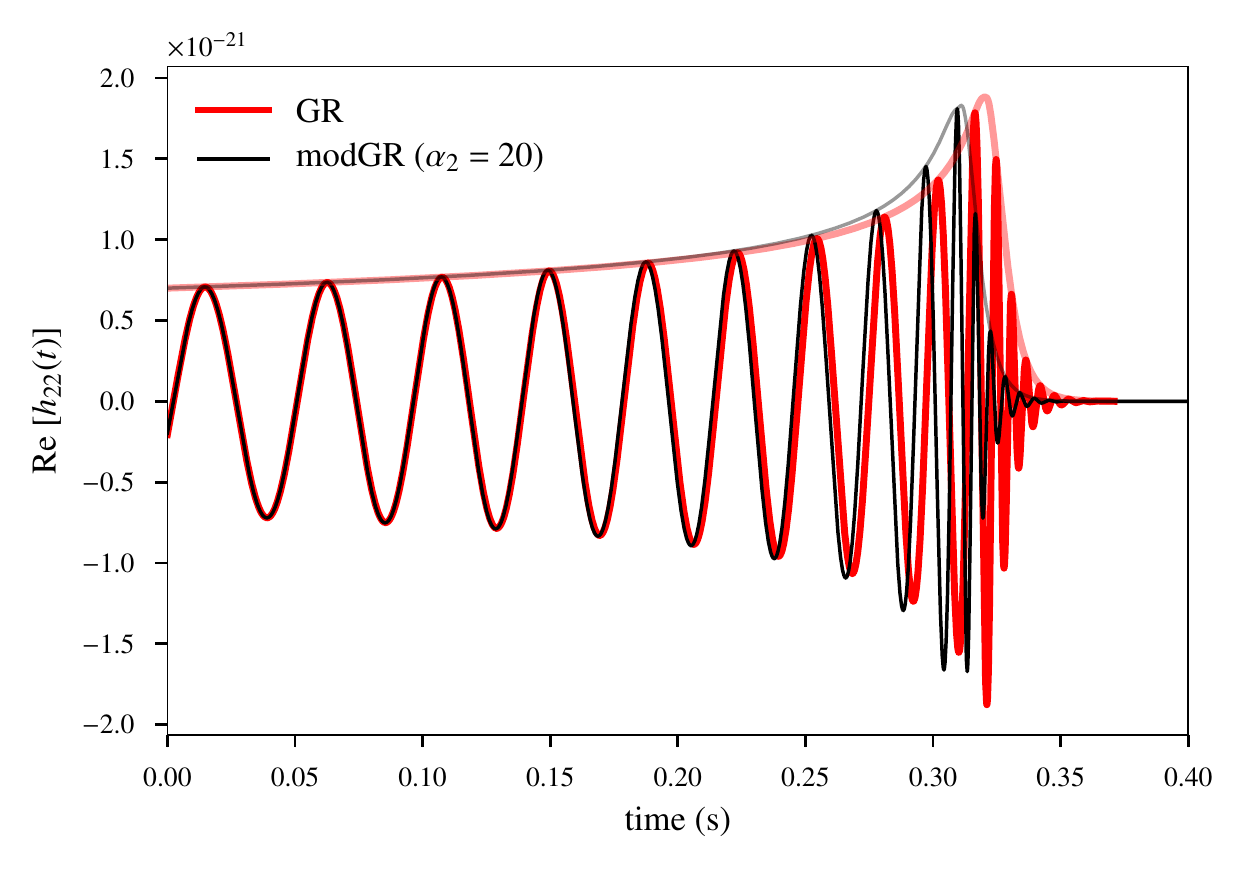}
\caption{The real part (darker lines) and amplitude (lighter lines) of the $(2,2)$ mode gravitational waveform form an equal-mass nonspinning binary black hole computed using the IHES EOB code with no modification (GR) and with our modification to the flux with $\alpha_2 = 20$ (modGR). For this illustration, we have taken the total mass to be $100M_\odot$, the distance to be $1~\mathrm{Gpc}$, and have aligned the waveforms at $t = 0$, which we have taken to correspond to a $(2,2)$ mode frequency of $20~\mathrm{Hz}$.}
\label{fig:modGR_merger-ringdown}
\end{figure*}

We retain the termination condition for the inspiral--plunge phase of the waveform used the original code. This uses the maximum of the orbital frequency (calculated from the EOB Hamiltonian) to mark the termination of the calculation of the inspiral--plunge waveform (by evolving the EOB Hamilton equations) and the start of the matching to the QNMs used to model the merger and ringdown. This termination condition coincides with crossing the light ring in the extreme mass-ratio limit. We also keep the next-to-quasicircular parameters set to the values determined from NR simulations that are already given in the code, for simplicity. Additionally, we keep all the other parameters set to their default values, except for a parameter that keeps one from probing unphysical regions of the EOB potential while the EOB equations are solved numerically. This was necessary to change in order to compute waveforms for higher mass ratios, even without other modifications to the code.

With these choices, we find that the waveforms look similar to a GR waveform up to the values of $\alpha_2 = 20$ we consider here. This is illustrated in figure~\ref{fig:modGR_merger-ringdown} for an equal-mass binary; the unequal-mass waveforms we consider in the next subsection also look similar to GR waveforms. Additionally, we have checked that this modification to the waveform does not induce any excess eccentricity: The eccentricity of the modified waveforms is small, and the same size as the eccentricities of the unmodified waveforms ($\lesssim 10^{-5}$ for the initial frequencies we are considering).\footnote{We estimate the eccentricity of the waveforms as in section~IV~C of~\cite{Dietrich:2015pxa} [Eq.~(4.13) and following], though with two additional higher-order PN terms added to the fitting function, since the eccentricity is small. The results did not change upon adding a third additional PN term. Since this estimate only uses the leading-order GW effects, it does not need to be modified to account for the modification to the flux, which only starts at $2$PN.}

\begin{figure*}[tbh]
\centering
\subfloat{
\includegraphics[width=0.5\textwidth]{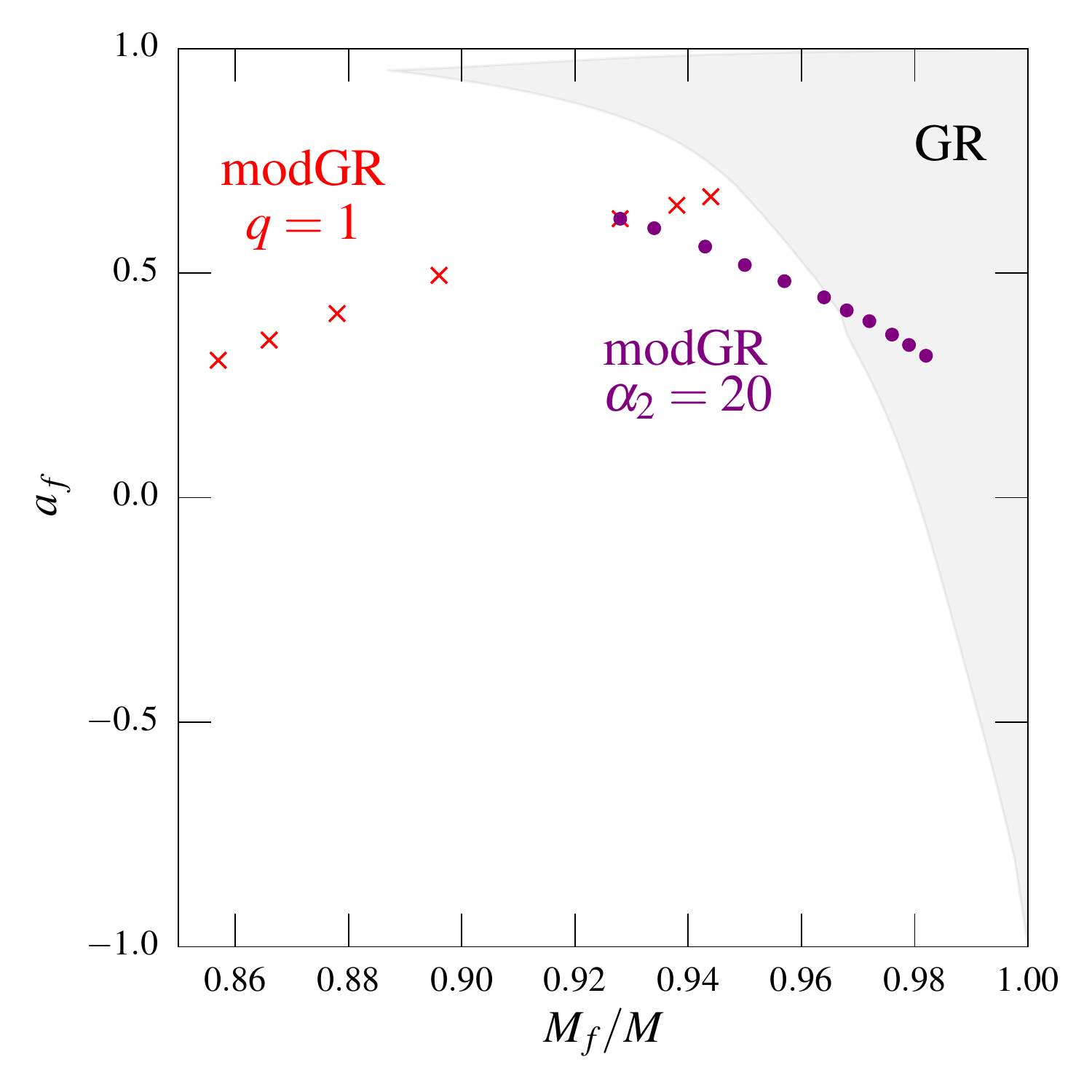}
}
\subfloat{
\includegraphics[width=0.5\textwidth]{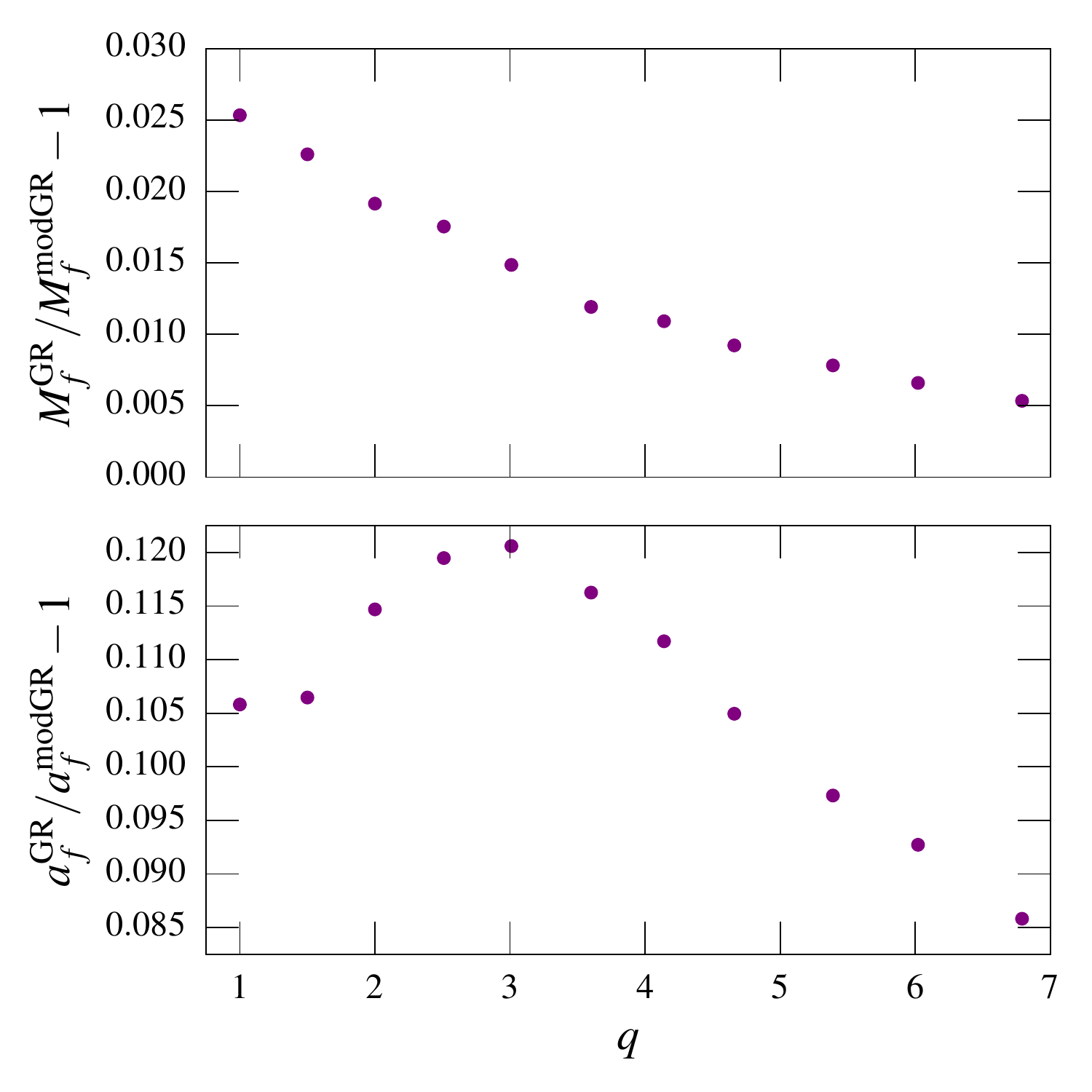}
}
\caption{\emph{Left:} The final mass and spin allowed by GR (shaded region), as calculated from the fits from~\cite{Healy:2014yta}, as well as the values for the kludge waveforms with a mass ratio of $q = 1$ and $\alpha_2 \in \{5, 10, 20, 100, 200, 300, 400\}$ (crosses, with $\alpha_2$ increasing from right to left), as well as $\alpha_2 = 20$ and $q$ from $1$ to $6.8$ in roughly steps of $0.5$ (circles, with $q$ increasing from left to right). The $\alpha_2 = 20$ points are taken from the injection set. \emph{Right:} The fractional differences between the final mass and spin for the kludge waveforms with $\alpha_2 = 20$, as a function of the mass ratio $q$.}
\label{fig:modGR_Mf_af}
\end{figure*}

Since the final mass and spin in the original EOB waveform are set by a fit to NR results, we replace this determination by demanding self-consistency of the radiated energy and angular momentum. That is, we choose the final mass and spin by minimizing the difference between the final mass and spin used to obtain the merger--ringdown part of the model and the final mass and spin obtained from energy and angular momentum balance using the initial energy and angular momentum and the radiated energy and angular momentum (calculated from the waveform). Here we are able to obtain fractional disagreements of less than $\sim 10^{-4}$ between the final mass and spin calculated from the radiated energy and angular momentum and those used to compute the merger-ringdown part of the signal. These fractional disagreements are considerably better than the agreement of these quantities for the original IHES EOB waveform for some mass ratios, and much smaller than the differences in radiated energy and angular momentum between the different waveforms we consider. We compute the energy and angular momentum loss using multipoles up to $\ell = 7$, since we are only able to fit for the ringdown up to $\ell = 7$, the highest $\ell$ for which the tabulated QNM results used in the code are available~\cite{Berti:2009kk, Berti_QNM_fits_url}. We take the standard GR expressions for the radiated energy and angular momentum to still be valid for this modified gravity waveform; these expressions indeed remain valid for a significant number of modified theories~\cite{Stein:2010pn}. We find that the final mass and spin falls outside the region allowed in GR in many cases, though there is a region of overlap for $\alpha_2 = 20$ and mass ratio $q := m_1/m_2 \gtrsim 4$; see figure~\ref{fig:modGR_Mf_af}. We also plot the fractional difference in the final mass and spin between GR and the kludge waveforms with $\alpha_2 = 20$ in this figure, and see that the fractional difference in the final spin is significantly larger than the fractional difference in the final mass.

In addition to enhanced energy and angular momentum fluxes, these kludge modified GR waveforms have an enhanced linear momentum flux. This leads to larger GW recoils than one obtains for nonspinning systems in GR, where the maximum is only $\sim 180~\mathrm{km\,s^{-1}}$ (see, e.g., \cite{Healy:2016lce}). For the $\alpha_2 = 20$ deviation we consider here, the maximum recoil is $\sim 440~\mathrm{km\,s^{-1}}$.\footnote{The original IHES EOB waveform model (without deviations) already has a maximum recoil of $\sim 350~\mathrm{km\,s^{-1}}$.} We do not include any potential effects (e.g., the Doppler shifts discussed in~\cite{Gerosa:2016vip}) from these enhanced recoils, which are still small (at most $\sim 10^{-3}$ times the speed of light).\footnote{Enhanced recoils were not an issue for the larger deviation considered in \paperone because the recoil vanishes, by symmetry, for the equal-mass nonspinning system considered there.}

We have not changed the QNM spectrum of the final black hole in these kludge modified gravity waveforms for simplicity. However, one expects the QNM spectrum to be modified in any alternative theory that predicts a different GW energy loss for a binary black hole, even those for which the unperturbed Kerr metric is a solution of the field equations, since the behavior of perturbations of Kerr will generically be different in an alternative theory (see, e.g., \cite{Barausse:2008xv}). Extending this analysis to the case where one modifies the QNM spectrum would be a useful additional check of the consistency test presented here, though the requisite catalogs of QNMs are not yet available, even for the case of a Kerr--Newman black hole, which could be used as a stand-in for a modified gravity black hole: See, however, Dias \emph{et al}.~\cite{Dias:2015wqa} for some recent results that could be used to obtain such a QNM catalogue.

We only inject the $(2,\pm 2)$ modes of the waveform here, since these are the only modes included in the waveform model we use for the test, as well as for the injections in section~\ref{sec:GRsims}. The modifications to the higher modes appear in the $(2,\pm 2)$ modes by their effect on the waveform's phasing, since the higher modes are used to compute the modified energy flux. We have run the test on two GR injections with higher modes in section~\ref{sec:precHM}, and found that the inclusion of higher modes does not bias the test in those cases.

\begin{figure}
\centering 
\includegraphics[height=2.5in]{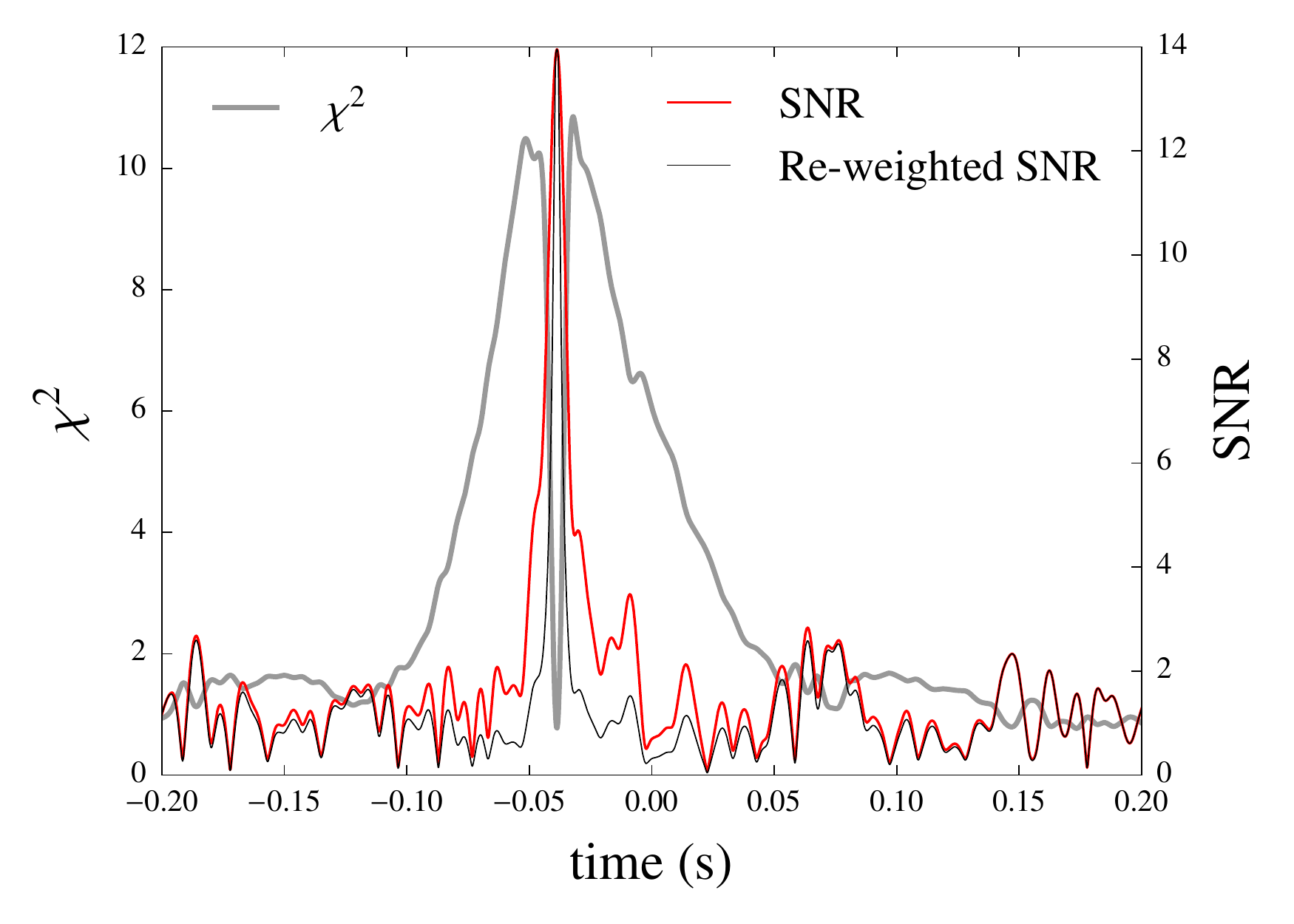}
\caption{Time series of the matched filter SNR, the reduced $\chi^2$ and re-weighted SNR from an injected modified GR waveform, corresponding to the second entry in table~\ref{tab:pycbc_results_modgr_inj}. The re-weighted SNR is the PyCBC detection statistic and is defined in equation~(3) of~\cite{Usman:2015kfa}. The injected signal is located in the data at around time $-0.05~\mathrm{s}$ in the plot. This is very similar to the analogous plot for GW150914 given in figure~8 of~\cite{TheLIGOScientific:2016qqj}, indicating that this modified GR signal would be found by the PyCBC search in a similar way to a genuine GR signal.}
\label{fig:pycbc_results_modgr_inj}
\end{figure}

Finally, almost all of our modified GR signals will be confidently detected by the standard matched-filter based searches for binary black holes, which include the chi-squared discriminatory test~\cite{Allen:2004gu}. To demonstrate this, we have run the PyCBC~\cite{Usman:2015kfa} matched filter based detection pipeline on these injections (using a single GR template from the \textsc{SEOBNRv2\_ROM\_DoubleSpin} family, corresponding to the parameters of the simulated modGR signal). We see that the chi-squared weighted SNR recovered by the detection pipeline is close to the optimal SNR of the signal (see table~\ref{tab:pycbc_results_modgr_inj} and figure~\ref{fig:pycbc_results_modgr_inj} for some examples). The SNR maximized over the full template bank is expected to be even larger than this, practically ensuring confident detections. 

\begin{table}
\centering
\begin{tabular}{c@{\quad}c@{\quad}c@{\quad}c@{\quad}c@{\quad}c@{\quad}c}
\toprule
$m_1~(M_\odot)$ & $m_2~(M_\odot)$ & $D_\mathrm{L}$ (Mpc) & Optimal SNR & PyCBC SNR & Reduced $\chi^2$ & Re-weighted SNR  \\
\midrule
139.1 & 69.5 & 5625.8 & 8.1 & 8.6 & 0.76 & 8.6 \\ 
88.1 & 14.7 & 803.5 & 14.3 & 14.0 & 0.78 & 14.0 \\ 
40.8 & 46.8 & 787.0 & 21.6 & 21.7 & 0.81 & 21.7 \\ 
\bottomrule
\end{tabular} 
\caption{The table shows the parameters of some of the injected modified GR signals ($D_\mathrm{L}$ is the luminosity distance) and how well the PyCBC pipeline is able to detect these injected signals in a single detector search (LIGO Livingston). Optimal SNR is the optimal SNR available in the injection, PyCBC SNR is the matched filter SNR extracted by the PyCBC pipeline, and re-weighted SNR is the chi-squared weighted SNR defined in equation~(6) of~\cite{Usman:2015kfa}. The optimal SNR is computed using the detector's noise power spectrum (it is defined by $\langle h | h \rangle^{1/2}$, for a signal $h$), while the matched filter SNR is a statistical variable computed using a specific noise realization, which is why the PyCBC SNR can be larger than the optimal SNR.}
\label{tab:pycbc_results_modgr_inj}
\end{table}

\subsection{Results from the simulations of modified GR signals}
\label{sec:modgr_results}

\begin{figure*}[tbh]
\centering
\includegraphics[height=2.1in]{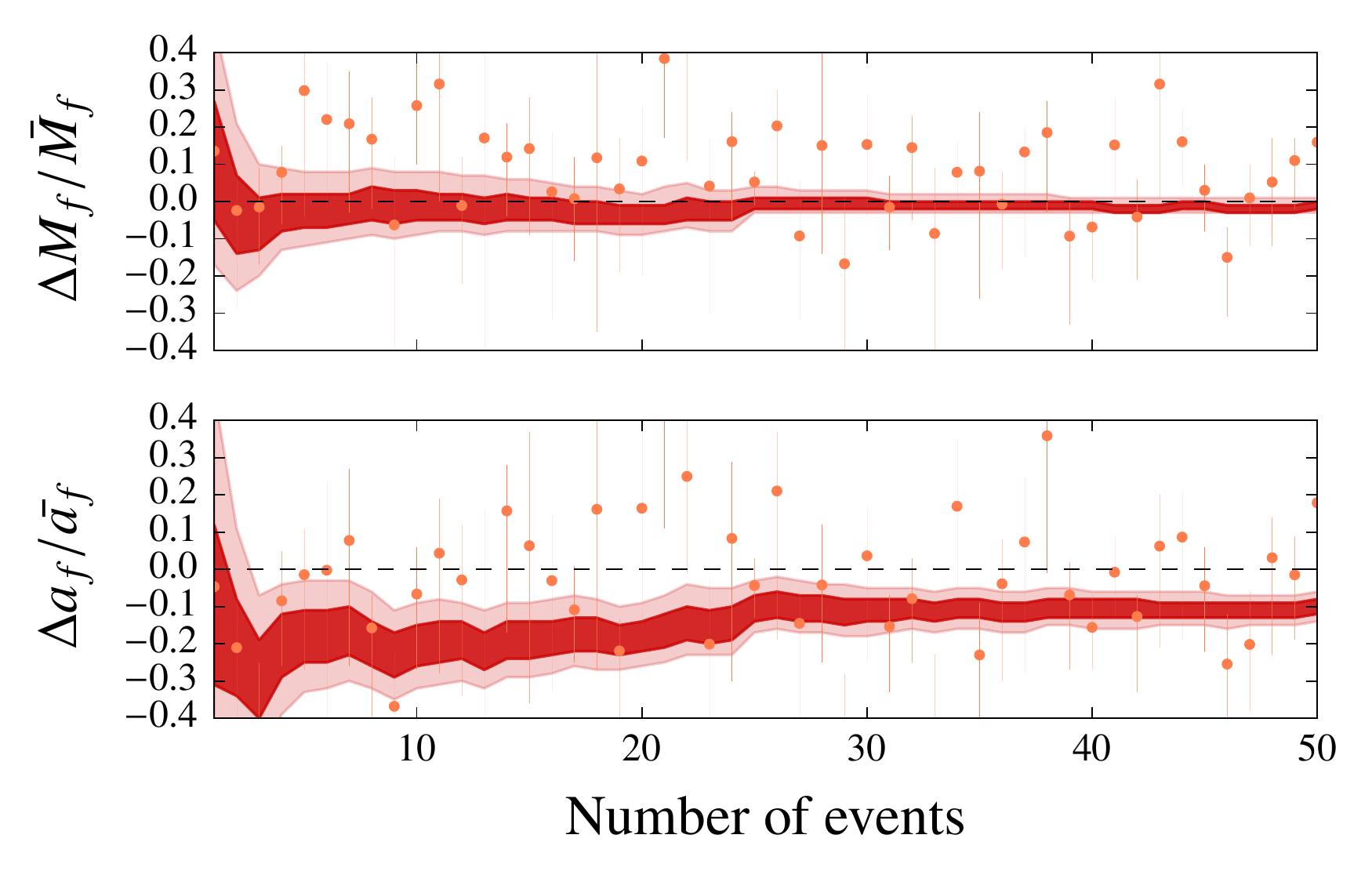}
\includegraphics[height=2.1in]{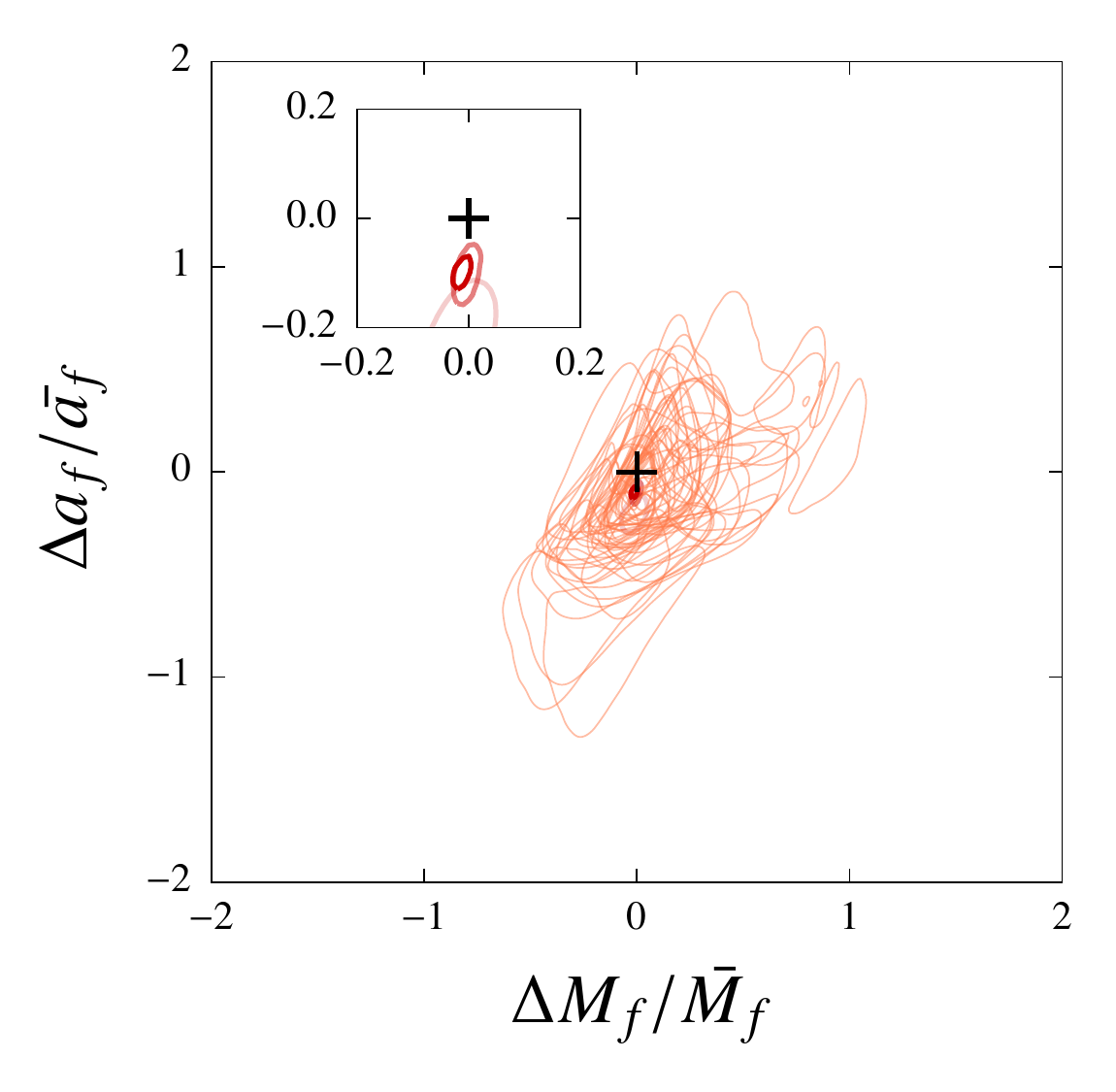}
\includegraphics[height=2.1in]{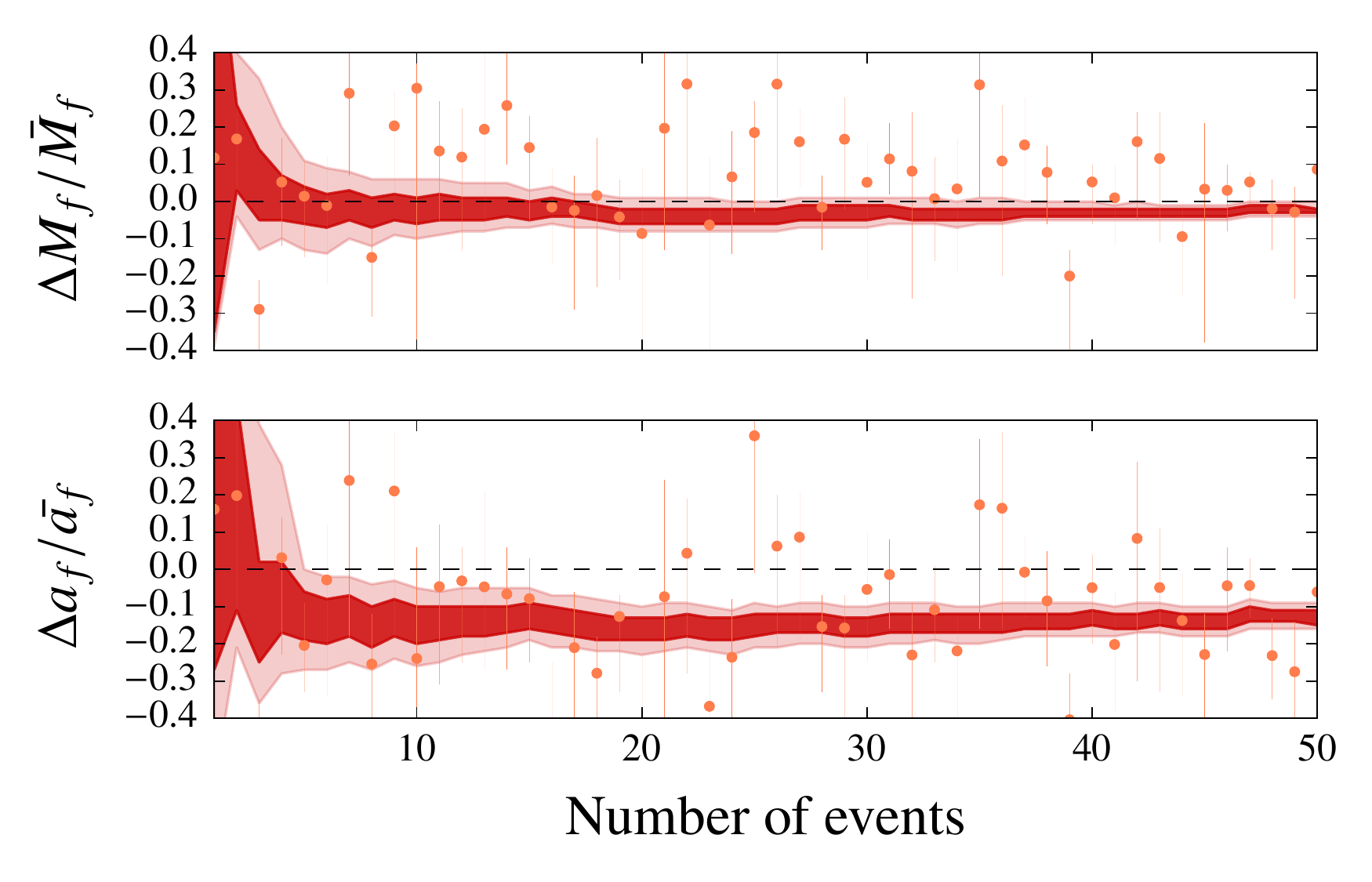}
\includegraphics[height=2.1in]{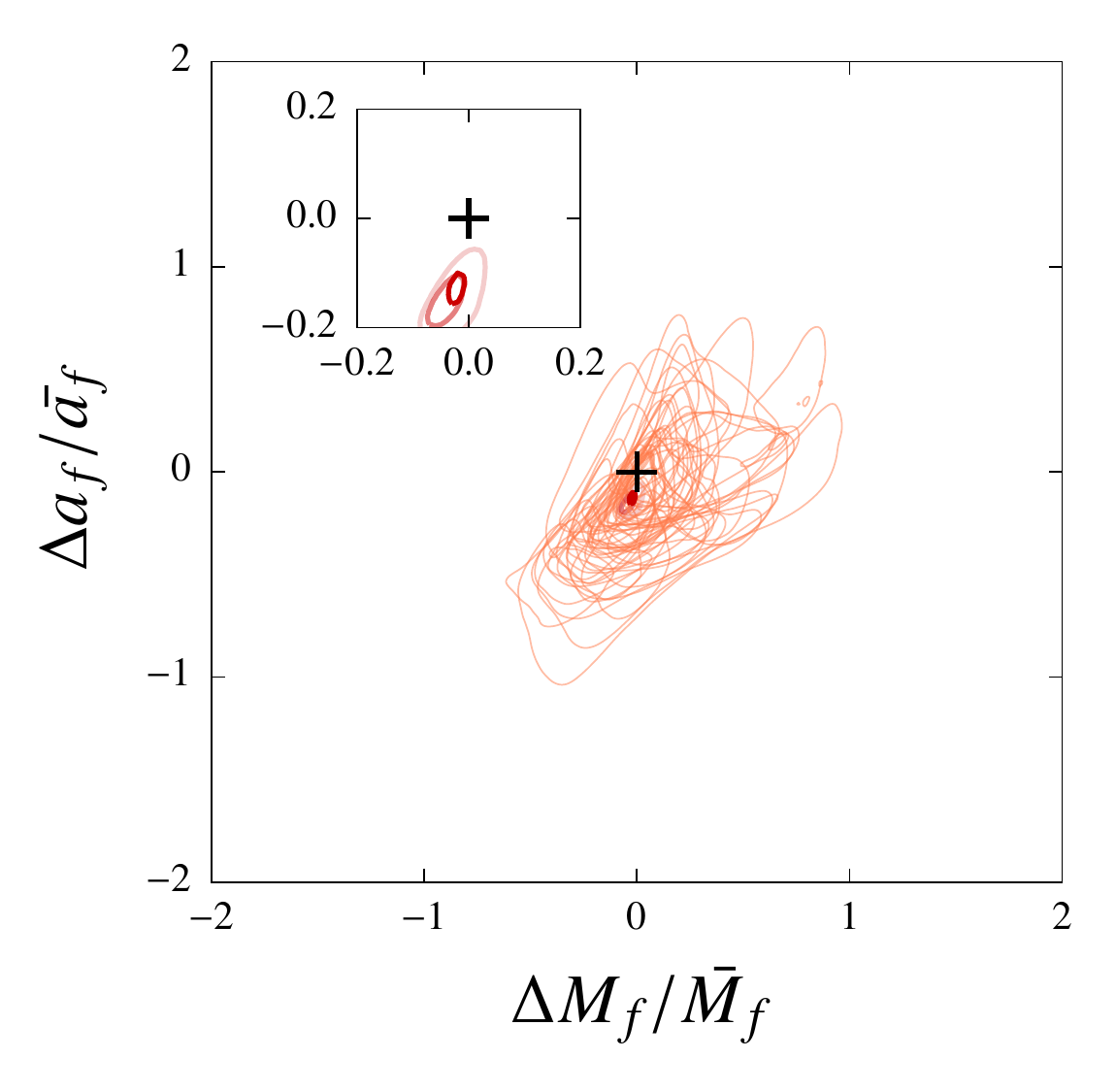}
\includegraphics[height=2.1in]{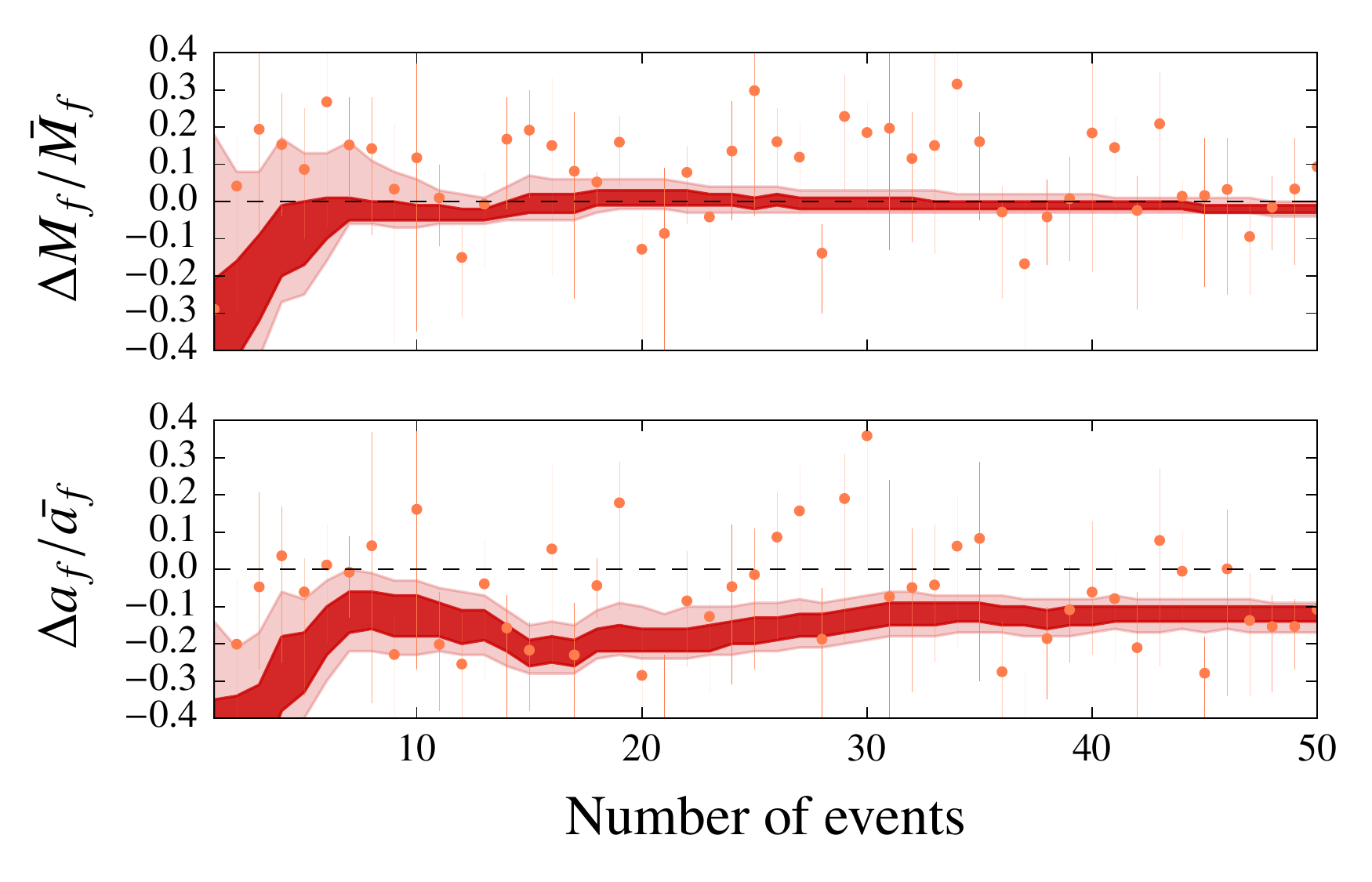}
\includegraphics[height=2.1in]{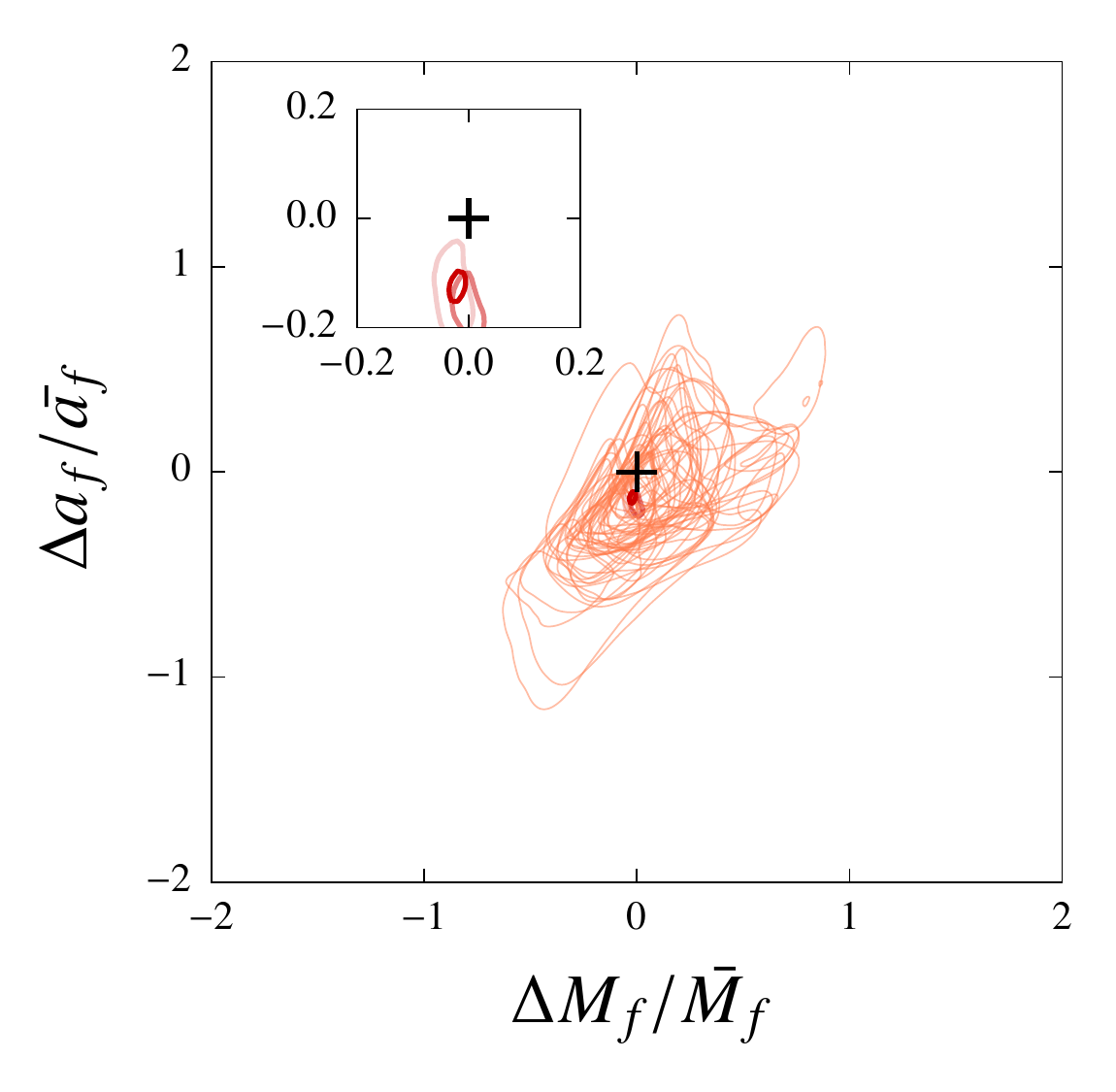}
\caption{Same as figure~\ref{fig:jointpost_gr} except that the test is performed on simulated signals containing a modification from GR described in section~\ref{sec:modgr_results}. The combined posteriors from multiple observations show a clear departure from the GR predictions (horizontal dashed lines on the left plots and the plus sign on the right plots).} 
\label{fig:jointpost_modgr}
\end{figure*}

Here we demonstrate that the IMR consistency test is able to identify (at least certain types of) deviations from GR by performing our analysis on a population of simulated signals using the modified GR waveforms described above. In \paperone, we have demonstrated the ability of this test to identify a relatively large modification to the binary's energy flux ($\alpha_2 = 400$). Such a deviation from GR was easily detectable with high confidence from a single observation of moderate SNR. Here we consider the ability of the test to discern much smaller deviations from GR by combining results from multiple observations. Specifically, we consider the same population of binary parameters considered to simulate the GR case in section~\ref{sec:GRsims} except that the waveforms are generated with a modification of the GR energy flux as described in section~\ref{sec:mod_gr_waveforms}. Also, for simplicity, we consider binary black holes with zero spins, since the EOB waveform family that we employ to produce modified GR waveforms is a nonspinning model~\cite{Damour:2012ky, IHES_EOB_url}. However, we still perform parameter estimation using the same \textsc{SEOBNRv2\_ROM\_DoubleSpin} aligned-spin model employed in section~\ref{sec:GRsims}.

Modified GR waveforms for binaries with different mass ratios (as well as distances, sky locations, and other extrinsic parameters) were constructed using the prescription presented in section~\ref{sec:mod_gr_waveforms}, with deviation parameter $\alpha_2 = 20$. An example of a modified GR waveform (along with the corresponding GR waveform from the same binary) is shown in figure~\ref{fig:modGR_merger-ringdown}. Although such a modified GR waveforms appears similar to a GR waveform, we show below that the IMR consistency test is able to identify deviations from GR by combining multiple events. Figure~\ref{fig:jointpost_modgr} shows the joint posteriors on the parameters that describe deviations from GR. It can be seen that, although individual posteriors are unable to identify deviations from GR, the combined posteriors show a clear departure from the GR predictions, even for the relatively small deviation that we consider.\footnote{Taking $\alpha_2 = 20$ corresponds to multiplying the $2$PN term in the frequency domain phase expression by a factor of $\sim 0.4$ for equal masses, with the factor decreasing monotonically to $\sim -4.5$ for a mass ratio of $7$ (the largest mass ratio in the detected population).} We also see that the combined posterior on $\sigma := \dafbyaf$  differs significantly more from $0$ than does the combined posterior on $\epsilon := \dMfbyMf$. This might be expected, due to the larger fractional differences from GR in the final spin seen in figure~\ref{fig:modGR_Mf_af}.


\subsection{Caveats}
\label{sec:caveats}

The way that we combine the posteriors from multiple events implicitly assumes that $\{\epsilon, \sigma\}$ take the same values in multiple events. This is certainly true if the true theory is GR, when they are always zero, but is not necessarily true for all modified theories of gravity. If the $\{\epsilon, \sigma\}$ values induced by the true theory are strongly dependent on the intrinsic parameters of the black holes (such as their masses and spins), then the combined posteriors from multiple events \emph{may} not converge to a non-zero value. Hence, this test (like any other test of GR) is not sensitive to \emph{all} kinds of possible deviations from GR. However, if the deviation is only weakly dependent on the intrinsic parameters, it is a reasonable assumption that the values of $\{\epsilon, \sigma\}$ observed by LIGO-Virgo will be clustered in a small range, due to the rather narrow range of masses ($\sim 10$--$80 M_\odot$) of black hole binaries employed in this test (and hence the reasonably narrow range of length scales probed by these black holes).\footnote{One might also worry about the spin-dependence of possible deviations. However, if the size of the deviation is controlled by the maximum curvature outside the horizons of the black holes, the spin only changes the length scale given by the curvature at the horizon of a Kerr black hole by a factor of $2\sqrt{2} \simeq 2.8$; see, e.g., Eq.~(5.47) of~\cite{2004rtmb.book.....P} for the expression for the Kretschmann scalar in Kerr, whose inverse fourth root gives the length scale we are considering. This neglects possible propagation effects, though even in that case LIGO is not expected to detect binaries with orders of magnitude different distances.} This is the case for simulations presented in section~\ref{sec:modGRsims}. Here the modification from GR has a dependence on the intrinsic parameters (see, e.g., figure~\ref{fig:modGR_Mf_af}); still, $\{\epsilon, \sigma\}$ estimates from individual posteriors tend to cluster around a reasonably small range of values, and the combined posteriors converge to a non-zero value in figure~\ref{fig:jointpost_modgr}. On the other hand, if the deviation is a strong function of the intrinsic parameters, then combining the posteriors of the entire population may not be the most sensitive method of detecting the deviation. One possibility to address this would be to combine the posteriors of subsets of detections with similar inferred parameters. Such combined posteriors from different regions of the parameter space (e.g., low-mass or high-mass) will provide us some indication of the nature of the deviation as a function of some scales (e.g., curvature). We leave such studies as future work.

Note that, in Figure~\ref{fig:jointpost_modgr}, the fractional difference in the final spin has the opposite sign to what might na{\"\i}vely be expected, given that the modGR waveforms correspond to smaller final spins than their GR counterparts. Since most of the angular momentum loss comes during the merger and ringdown, it seems reasonable that the inspiral values for the final mass and spin would be close to their GR values for the system's mass ratio, while those inferred from the ringdown would correspond to the true values used to construct the waveform. However, such an interpretation is possible only when the parameters are estimated from only the \emph{early} inspiral and \emph{late} ringdown parts of the waveform. Such a test would require both the early inspiral and late ringdown to be observed with high SNR. While such tests might be possible in the future, currently we are limited by the moderate SNRs of the observed signals, and hence are forced to include the late inspiral and merger parts in our analysis. For the case of the modGR waveforms that we use, the estimated parameters from the inspiral part are biased (typically to more massive, asymmetric binaries) which correspond to a smaller value of the final spin, as compared to the same estimated from the post-inspiral part, producing the observed sign in the fractional difference in the final spin. 

\section{Robustness of the consistency test}
\label{sec:robustness}
We also test the robustness of the IMR consistency test to the choice of various parameters used in the analysis, for example, the cutoff frequency used to demarcate the division between the low-frequency (inspiral) and high-frequency (merger--ringdown) parts of the waveform, the choice of the particular approximant waveform, and the fitting formula for the mass and spin of the final black hole, etc. For this we use a simulated GR signal with waveform approximant \textsc{SEOBNRv2\_ROM\_DoubleSpin} (except in section~\ref{sec:robustNRFit} where we use approximant \textsc{IMRPhenomPv2}~\cite{Hannam:2013oca,Khan:2015jqa,Boheetal_IMRPhenomPv2} and in section~\ref{sec:precHM} where we use NR waveforms). We also study the effect of neglecting higher modes and spin precession on the IMR consistency test. These are performed by injecting NR waveforms from the publicly available Simulating eXtreme Spacetimes (SXS) catalog~\cite{SXS-Catalog}. All these studies are performed assuming a binary black hole signal with parameters close to that of the first LIGO event GW150914 --- masses $m_1 = 36 \msun$, $m_2 = 29 \msun$ and aligned spins, $a_{1,z} = -0.32$, $a_{2, z} = 0.58$, with optimum sky position and orientation, producing an optimal SNR of $25$ in the Advanced LIGO Hanford--Livingston and Advanced Virgo network. These studies demonstrate the robustness of the IMR consistency test only for the case of signals with parameters similar to that of GW150914, that is, with moderate mass ratios. If we observe signals with large mass ratios and large, misaligned spins in the future, the robustness of the IMR consistency test will need to be reinvestigated. Also, these robustness studies have been restricted to the case of single events: when a large number of events are combined to produce precise constraints on the deviations from GR, we will have to worry about even small systematic errors affecting our analysis. Such studies have to be performed in the near future, in anticipation of the large number of binary black hole signals that Advanced LIGO is expected to observe. 

\subsection{Cutoff frequency between the inspiral and merger--ringdown}
\label{sec:robustCutoff} 
There is no well defined transition frequency (or time) between the inspiral and post-inspiral (merger--ringdown) parts of the waveform~\cite{Buonanno:2006ui}. We have chosen the ISCO frequency of the final Kerr black hole with mass and spin inferred from the full IMR waveform at this cutoff frequency. Increasing this cutoff frequency will increase (decrease) the SNR of the inspiral (post-inspiral) part and hence will improve (worsen) the parameter estimation from the inspiral (post-inspiral) analysis, thus affecting our statistical errors. There thus is likely an optimal choice for the cutoff frequency as far as the test's sensitivity to a given deviation from GR is concerned. However, if our analysis is free from major systematic errors, the two independent estimates always have to be consistent with each other when it is applied to a binary black hole coalescence in GR. Here we illustrate that our analysis produces consistent results for reasonable choices of the cutoff frequency.   

The ISCO frequency of a Kerr black hole with mass $M_f = 62 M_\odot$ and dimensionless spin $a_f = 0.68$ (remnant of the merger of two black holes with masses $m_1 = 36 \msun$, $m_2 = 29 \msun$ and aligned spins, $a_{1,z} = -0.32$, $a_{2, z} = 0.58$) is $146.4~\mathrm{Hz}$. We repeat the analysis with 4 different choices of the cutoff frequency in the interval $50$--$150~\mathrm{Hz}$. The results are shown in figure~\ref{fig:robustness_plots} (top panels). It can be seen that, while the spread of the posteriors (that is, the statistical errors) depends on the choice of the cutoff frequency, the inspiral and post-inspiral estimates are always consistent with each other. Also, the posteriors on the parameters describing deviations from GR are always consistent with zero, indicating the robustness of the test against the specific choice of the cutoff frequency. 

\subsection{Waveform approximant}
\label{sec:robustWaveform} 
The Bayesian inference described in section~\ref{sec:method} is performed by employing semi-analytical gravitational waveforms (computed by a combination of analytical and numerical relativity) as the GR model. The inherent assumption is that these are faithful representations of the actual signals produced by nature. However, depending on the particular methods used to construct these waveforms, there can be minor differences between different gravitational waveform families (approximants). Here, we demonstrate the robustness of the IMR consistency test employing two IMR waveform families, namely, the non-precessing spin reduced-order EOB model \textsc{SEOBNRv2\_ROM\_DoubleSpin}~\cite{Purrer:2015tud,Taracchini:2013rva} and the precessing single effective-spin phenomenological model \textsc{IMRPhenomPv2}~\cite{Hannam:2013oca,Khan:2015jqa,Boheetal_IMRPhenomPv2}. The systematic errors of these models in the vicinity of GW150914 and their effects on parameter estimation were studied in~\cite{Abbott:2016wiq}.

Figure~\ref{fig:robustness_plots} (second row) shows the results from simulated GR signals from a binary black hole with parameters described in the previous section. Apart from performing the analysis where the same waveform family is employed in estimating the parameters from the inspiral and post-inspiral parts, the figure also demonstrates the robustness of the consistency test when we employ one waveform family for estimating the parameters from the inspiral part and another to estimate the same from the post-inspiral part. In each case, the waveform family used to create a portion of the waveform, inspiral or merger-ringdown, is the same used to recover the parameters from that portion.

\subsection{Fit formulas for the mass and spin of the final black hole}
\label{sec:robustNRFit} 

The consistency test we have developed relies on NR fitting formulas to map the binary's initial masses and spins into the mass and spin of the remnant black hole. There are a number of such fits available in the literature, of varying degrees of accuracy and generality. Here we consider a variety of recent fits for the final mass and spin.\footnote{While there are fits for the final mass and spin used in the two waveform models we employ, we do not just use those here, since they are not the most accurate and up-to-date ones: see Jim{\'e}nez-Forteza \emph{et al}.~\cite{Jimenez-Forteza:2016oae}, which compares the final mass fit used in SEOBNRv2 and the final mass and spin fits used in IMRPhenomPv2 with more accurate fits. Note, however, that the additional calibrated terms in the merger-ringdown parts of these models may compensate for some of the slight inaccuracies of the final mass and spin fits used.} We consider one set that is only applicable to nonspinning binaries (so we do not use the estimated spins in applying this) given in equations~(29) of Pan~\emph{et al}.~\cite{Pan:2011gk}, and two sets applicable to binaries with aligned spins. The first set is from Healy, Lousto and Zlochower (HLZ)~\cite{Healy:2014yta} and involves reasonably complicated implicit expressions given in Eq.~(14) and Eq.~(16) in that paper, with coefficients given in Table~XI (we use the fourth-order fits). The second, simpler, though somewhat less accurate set is from Eq.~(3.6) and Eq.~(3.8) in Husa~\emph{et al}.~\cite{Husa:2015iqa} and only uses a single effective spin. The latter two fits were used in~\cite{Abbott:2016blz, TheLIGOScientific:2016wfe} to infer the final mass and spin of GW150914 and the HLZ fits were used in the implementation of the current test in \paperone and \cite{TheLIGOScientific:2016src}. For these fits, we use the projection of the spins along the orbital angular momentum when using a precessing waveform model, e.g., IMRPhenomPv2. The final mass and spin used to determine the ISCO frequency is not changed throughout the analysis, even when we are changing the fitting formulas. The calculation of the ISCO frequency uses the HLZ fits.

It is also possible to augment aligned-spin fits for the final spin to include the contribution from the in-plane spins, as was done for the HLZ fit in~\cite{LIGO-T1600168} and significantly increases its accuracy for precessing systems. As that reference shows for the HLZ fit, the basic aligned-spin final mass fit is already accurate for precessing systems when evaluated using the components of the spins along the orbital angular momentum. This augmentation was applied to the HLZ fit to infer the final spin of the binary black hole events from Advanced LIGO's first observing run in~\cite{TheLIGOScientific:2016pea,Abbott:2016nmj,Abbott:2016izl}. We thus also consider this extension of the final spin expression for the HLZ and Husa~\emph{et al}.\ fits (HLZ aug and Husa~\emph{et al}.\ aug), as well as a new final spin fit for precessing systems from Hofmann, Barausse and Rezzolla (HBR)~\cite{Hofmann:2016yih} (we use the one with the smallest number of coefficients, which the authors recommend), which we pair with the HLZ final mass fit. We also consider the augmented versions of both new aligned-spin fits that extend those in Husa~\emph{et al}.\ to include the spin difference (UIB aug)~\cite{Jimenez-Forteza:2016oae}, as well as a small update to the HLZ fits from Healy and Lousto (HL aug)~\cite{Healy:2016lce}. An average of these fits (HBR, UIB aug, and HL aug for the final spin, and just the second two for the final mass) was used to infer the final mass and spin of the binary coalescence that produced GW170104, as well as in the application of the IMR consistency test to that system~\cite{Abbott:2017vtc}.

In figure~\ref{fig:robustness_plots} (third row), we show that for the simple test case we consider, the two aligned-spin fits give almost identical results, as do the extensions using the in-plane spins, while the nonspinning fits still give a posterior that is consistent with zero. Here we use the precessing waveform model IMRPhenomPv2~\cite{Hannam:2013oca,Khan:2015jqa,Boheetal_IMRPhenomPv2} for both the injection and recovery. We use it for the recovery in order to have nonzero in-plane spins, and use it (in its aligned-spin limit) for the injection so we are only testing different fits, not different waveform models here. Since we apply the same fitting formulas to infer the final mass and spin from both the inspiral and post-inspiral portions of the waveform, systematic errors in the fitting formulas do not translate into systematic errors in the test. Nevertheless, having more accurate fitting formulas might help the test to identify modified GR signals.

\subsection{Effects of spin precession and higher modes}
\label{sec:precHM}

Although GW emission in the leading order is quadrupolar, higher order (nonquadrupole) modes can contain nonnegligible power if the system has large asymmetries (e.g., highly unequal masses or large, misaligned spins). Due the unavailability of waveform models that are accurate and computationally fast to generate, currently the parameter estimation is mostly performed making use of waveform templates that only model the quadrupole ($\ell = 2, m = \pm2$) modes. This can potentially introduce systematic errors in the estimated parameters, which could be falsely taken for a deviation from GR. However, the systematic errors due to neglecting nonquadrupole modes are expected to be negligible for the case of binary black holes with moderate mass ratios and aligned/antialigned spins~\cite{Abbott:2016wiq,Varma:2016dnf}. In this paper's population study we considered binaries with spins aligned/antialigned with the orbital angular momentum so that there is no precession. However, fast-to-evaluate waveform templates that take into account the dominant precession effects are available~\cite{Hannam:2013oca,Khan:2015jqa,Boheetal_IMRPhenomPv2}, which have been used for performing the IMR consistency test on the LIGO events GW150914~\cite{TheLIGOScientific:2016src} and GW170104~\cite{Abbott:2017vtc}. 

\begin{figure}
\includegraphics[height=1.9in]{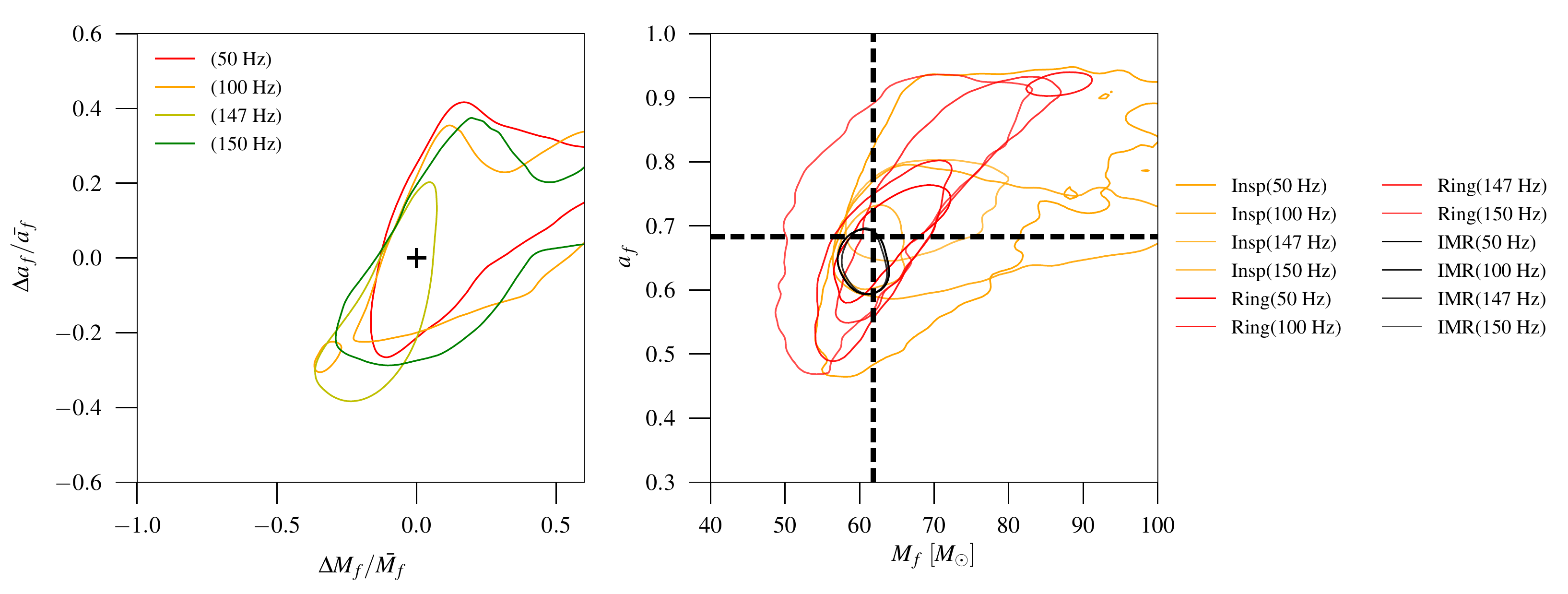}
\includegraphics[height=1.9in]{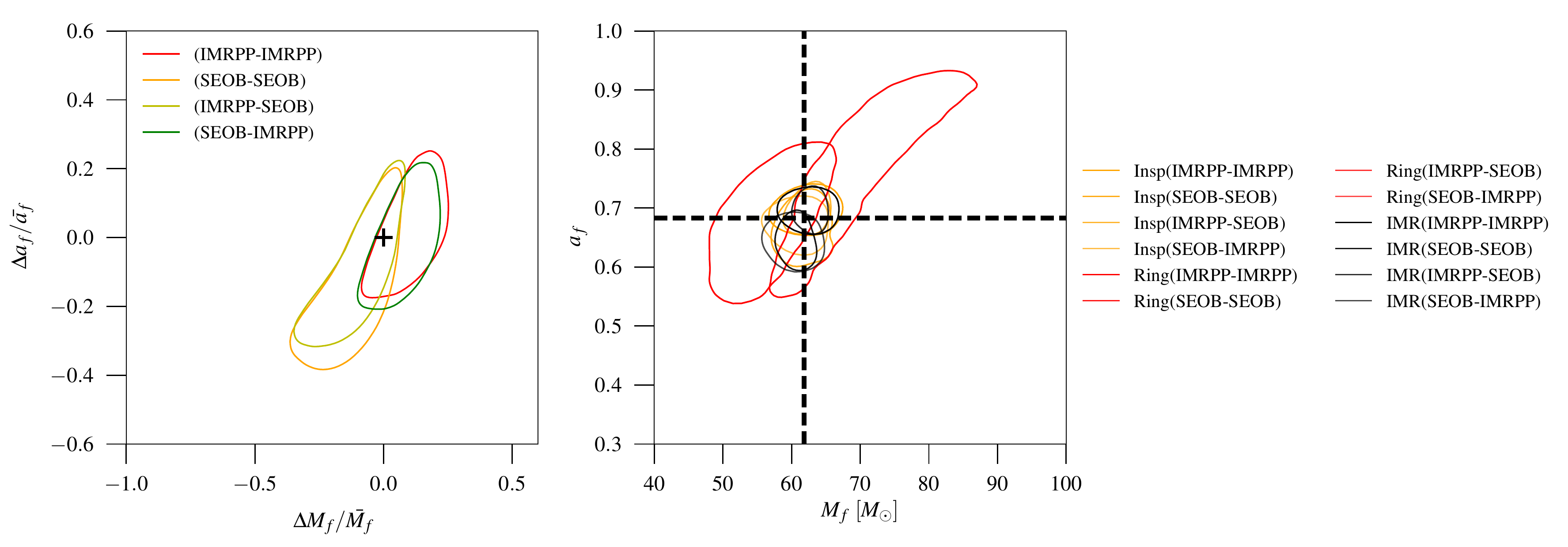}
\includegraphics[height=1.9in]{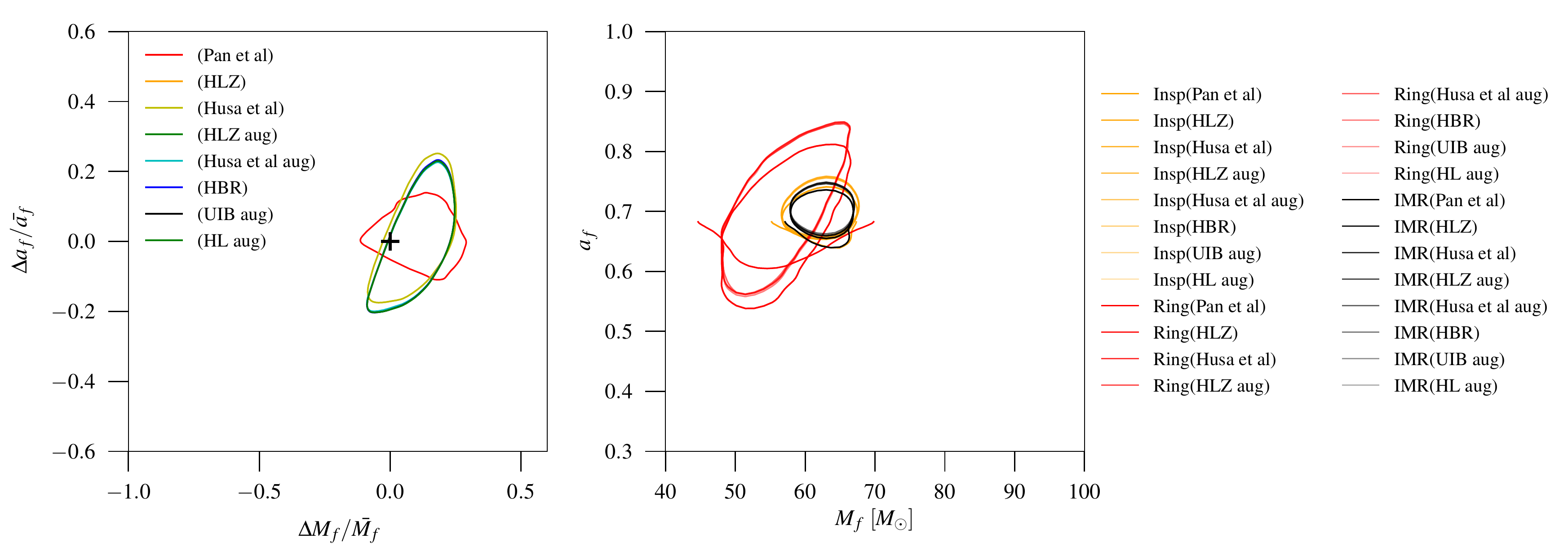}
\includegraphics[height=1.9in]{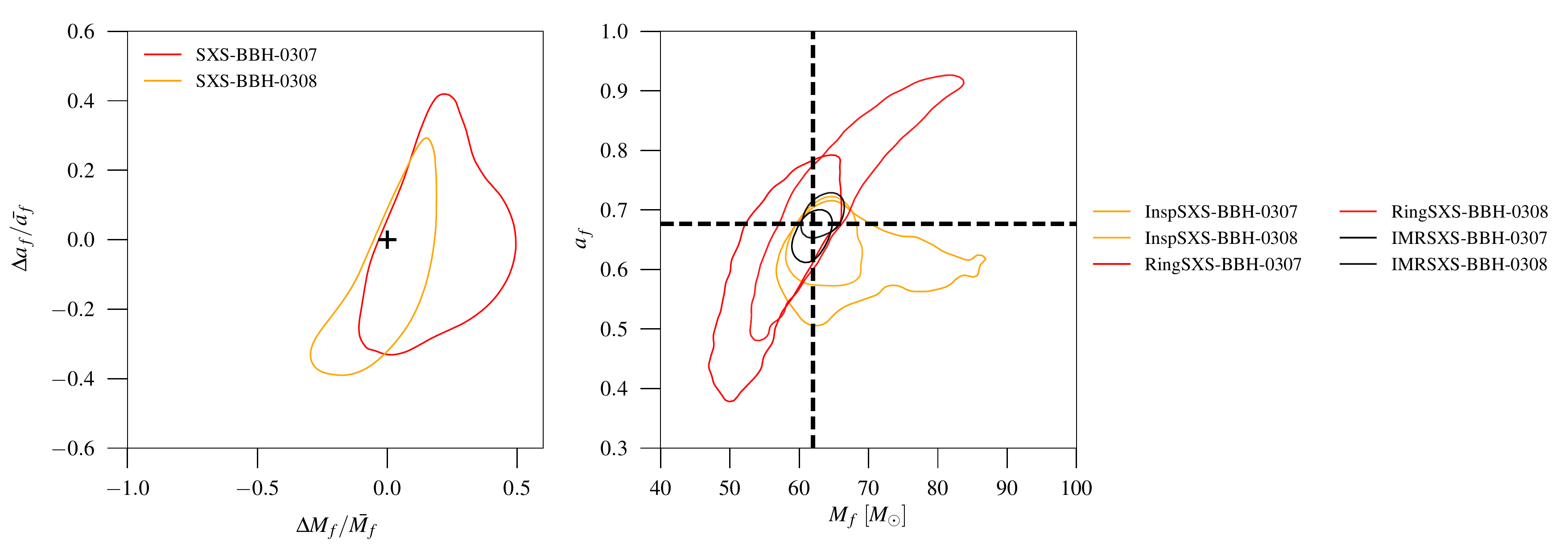}
\caption{Robustness studies against the choice of cutoff frequency (top row), approximants (second row), fitting formulas (third row) and effects of higher modes and precession (bottom row). Left panels show the $68\%$ credible region of the posterior distributions $P(\epsilon, \sigma)$. The GR value is marked by a + sign. Right panels show the $68\%$ credible region of the posterior distributions $\Pin(M_f,a_f)$, $\Prd(M_f,a_f)$ and $\Pimr(M_f, a_f)$ of the mass and spin of the final black hole estimated from the inspiral, merger--ringdown parts and the full IMR signal for each simulated case, respectively. In the second panel, the parentheses in the legend carry a combination of labels; the first label refers to the waveform family used to perform the injection and recovery of the inspiral portion, while the second label refers to the same for the merger-ringdown portion. In the third panel, the contours for the five fits that include the contributions from in-plane spins to the final spin (HLZ aug~\cite{Healy:2014yta}, Husa \emph{et al}.\ aug~\cite{Husa:2015iqa}, HBR~\cite{Hofmann:2016yih}, UIB aug~\cite{Jimenez-Forteza:2016oae} and HL aug~\cite{Healy:2016lce}; ``aug'' denotes augmentation following~\cite{LIGO-T1600168}) lie almost on top of each other, thus making the individual contours indistinguishable. The same is true for the contours for the purely aligned-spin fits (HLZ~\cite{Healy:2014yta} and Husa~\emph{et al}.~\cite{Husa:2015iqa}). The nonspinning Pan~\emph{et al}.~\cite{Pan:2011gk} fit has a distinct contour.}
\label{fig:robustness_plots}
\end{figure}

We investigate the robustness of the IMR consistency test against the effects of precession and higher modes, making use of injected NR waveforms from the SXS waveform catalog~\cite{Mroue:2013xna, SXS-Catalog, Schmidt:2017btt}. We select two unequal mass ($m_1/m_2=1.228$) waveforms used in~\cite{Abbott:2016wiq}: SXS:BBH:0307, which has aligned spins ($a_{1,z} = 0.32$, $a_{2, z} = -0.5798$\footnote{The aligned spin components for the NR injections used in section~\ref{sec:precHM} have signs opposite to the ones used in the rest of the section; this is not a typographical error.}) binary black hole system, and SXS:BBH:0308, which is a precessing system with spin magnitudes $a_1 = 0.3406$, $a_2 = 0.6696$ and aligned components $a_{1,z} = 0.3224$, $a_{2, z} = -0.5761$. We inject both systems with total masses of $65 M_{\odot}$ (so their individual masses are $36$ and $29 \msun$, the same as for the previous injections), and with inclination angles of $0$. For parameter estimation, we use \textsc{SEOBNRv2\_ROM\_DoubleSpin} as our template. The results are summarized in figure~\ref{fig:robustness_plots} (bottom panel): the test is robust against the presence of higher modes and spin precession in the comparable-mass regime. 

\subsection{Effect of splitting the signal in the frequency domain}
\label{sec:spec_leakage}

In the IMR consistency test, we check the consistency of the posteriors of the final mass and spin estimated from the early part of the signal (inspiral) with the same estimated from the late part of the signal (merger--ringdown). The split between the inspiral and merger--ringdown parts is done in the Fourier domain. This choice is purely made for convenience, since, in the Fourier domain, this amounts to simply setting the lower and upper limits of the likelihood integral in equations (\ref{eq:likelihood}) and (\ref{eq:nwip}).\footnote{This is different from splitting the signals in time domain and then taking the Fourier transform, which will introduce additional artifacts in the Fourier transform due to edge effects or windowing.} There is a possibility that a fraction of the power from the inspiral can get deposited at high Fourier frequencies, or the power from the merger--ringdown parts to get deposited at low Fourier frequencies. This effect will be particularly pronounced for the ringdown signal, whose power is spread over a large range of Fourier frequencies. One may worry that this spectral leakage would cause the posteriors derived from the two regions of Fourier frequency to be automatically consistent (although, we have already demonstrated that when certain kinds of deviations from GR are present in the signal, the IMR consistency test is able to identify them; see section~\ref{sec:modgr_results}).

Below we show that the effect of spectral leakage is small for the case of cutoff frequencies that we choose. In order to demonstrate this, we do the following: We compute the Fourier transform of the NR waveform SXS:BBH:0253 (an aligned-spin system with $m_1/m_2 = 2$ and $a_{1,z} = a_{2,z} = 0.5$)~\cite{Mroue:2013xna, SXS-Catalog} making use of the \emph{stationary phase approximation} (SPA) [e.g., equation~(3.7) of~\cite{Damour:2000gg}]. The approximation here is that the power at a Fourier frequency $f$ comes entirely from the time $t_f$ (the saddle point) when the instantaneous frequency $F(t) := (\dd \varphi(t)/\dd t)/2\pi$ is equal to $f$. Here $\varphi(t)$ is the phase of the waveform. In this approximation, the magnitude of the Fourier transform of the NR waveform can be computed as
\begin{equation}
|\tilde{h}(f)| = \frac{A(t_f)}{\sqrt{\dot{F}(t_f)}}
\end{equation}
where $A(t)$ is the time-domain amplitude of the NR waveform, and a dot denotes a time derivative. 

\begin{figure}
\centering 
\includegraphics[height=3in]{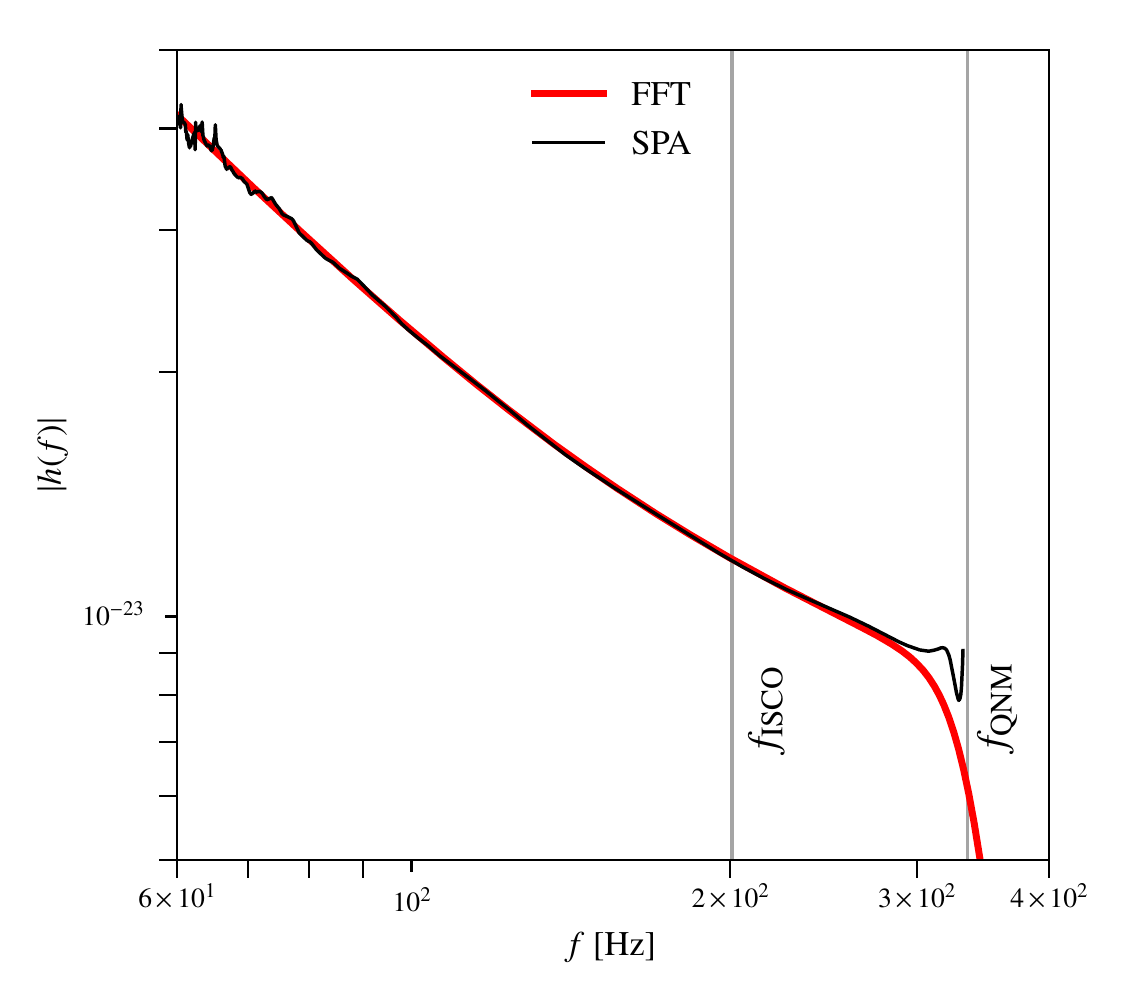}
\caption{Comparison of the Fourier transform of the NR waveform SXS:BBH:0253 computed using the SPA with the exact Fourier transform computed using the FFT. The corresponding ISCO frequency $f_\mathrm{ISCO}$ is also shown, which is used to demarcate the inspiral and merger--ringdown parts. The plot suggests that the power at Fourier frequencies less than $f_\mathrm{ISCO}$ can be fully explained to be coming from the early times of the waveform. Also shown is the dominant QNM frequency, for reference. At high frequencies, the agreement between SPA and FFT becomes poor, indicating spectral leakage. The noise in the SPA at lower frequencies is due to the numerical noise in computing the frequency derivative.}
\label{fig:spa_fft_comparison}
\end{figure}

If the Fourier transform computed using the SPA agrees well with numerically computed Fast Fourier Transform (FFT), this gives a strong indication that the spectral leakage is negligible. Figure~\ref{fig:spa_fft_comparison} compares the Fourier transform computed using the SPA with the FFT. The corresponding ISCO frequency $f_\mathrm{ISCO}$ is also shown, which is used to demarcate the inspiral and merger--ringdown parts. The plot suggests that the power at Fourier frequencies less than $f_\mathrm{ISCO}$ can be fully explained to be coming from the early times of the waveform (i.e., with instantaneous frequency $F(t) < f_\mathrm{ISCO}$). The excellent agreement between the SPA and FFT at frequencies less than $f_\mathrm{ISCO}$ suggests that there is no appreciable spectral leakage between the two bands --- almost all the power in the $f < f_\mathrm{ISCO}$ ($f > f_\mathrm{ISCO}$) band should come from the early (late) times. However, if we choose a transition frequency much higher than $f_\mathrm{ISCO}$ to demarcate the inspiral and merger--ringdown (say, the dominant QNM frequency, which is also shown in the figure) we expect significant spectral leakage between the two bands. Thus, if we need to perform this same analysis purely on the ringdown part (without including the merger), the analysis must be performed in the time domain. There is ongoing work in this direction. 

\section{Conclusions and future work}
\label{sec:conclusions}
In this paper we have provided a detailed description of a test of GR based on the consistency between the inspiral, merger and ringdown parts of observed GW signals from coalescing binary black holes. First presented in \cite{Ghosh:2016xx}, based on the idea proposed by Hughes and Menou~\cite{Hughes:2004vw}, this was one of the handful of tests used to establish the consistency of the GW150914 event with a binary black hole system described by GR~\cite{TheLIGOScientific:2016src}. Here, we demonstrate how the likelihoods from multiple binary black hole events can be combined to produce tighter constraints on parameters describing a deviation from GR. In order to illustrate how this test might be able to detect certain types of deviations from GR, we generate internally consistent kludge waveforms which disagree with the GR prediction of the radiated energy and angular momentum, by increasing the flux radiated into GWs. We then simulate a population of binaries modeled by these modified GR waveforms. By combining the results from multiple events, we demonstrate how this test might be able to detect small deviations from GR. For the case of single events with parameters similar to LIGO's first binary black hole observation, we also demonstrate the robustness of the consistency test against specific choices of various analysis parameters, such as the choice of the transition frequency used to demarcate the inspiral and merger--ringdown parts, the specific waveform approximant, fitting formulas for the mass and spin of the remnant black hole, etc. 

While we have investigated the reliability of our test to ensure its robustness, there remain further extensions of the test and these analyses to pursue. In particular, we have assumed that black holes in binaries have nonprecessing spins in the populations we simulate in this paper. We have also neglected the effect of nonquadrupole modes in the parameter estimation (although the fitting formulas for the remnant mass and spin include their contributions to the energy and angular momentum loss). While the effects of precession and nonquadrupole modes are expected to be negligible for the parameter estimation of comparable-mass binaries (see, e.g.,~\cite{Abbott:2016wiq,Varma:2016dnf}) their effects might be appreciable for binaries with large mass ratios and high spins. Waveform templates describing the spin precession through an effective spin parameter, notably the \textsc{IMRPhenomPv2} template~\cite{Hannam:2013oca,Khan:2015jqa,Boheetal_IMRPhenomPv2}, are already available and have been used in parameter estimation. \textsc{IMRPhenomPv2} has been employed in the application of this test to GW150914 in~\cite{TheLIGOScientific:2016src}, to GW170104 in~\cite{Abbott:2017vtc}, and in some of the robustness tests in section~\ref{sec:robustness}. More recently, waveform templates describing the full double-spin precession effects have also become available~\cite{Pan:2013rra,Babak:2016tgq}. These templates can be employed in parameter estimation when precession effects are expected to be apparent. Fitting formulas for the remnant mass and spin valid for the case of precessing initial spins are already available; some of them have been used in this paper itself. Waveform templates for binaries of spinning black holes that include the effect of nonquadrupole modes have started to become available~\cite{London:2017bcn}, and they may help to avoid possible systematic errors due to neglecting nonquadrupole modes. 

The robustness studies performed in this paper are restricted to the case of single events with modest SNR. When a large number ($\sim100$) of events are combined to produce precise constraints on deviations from GR, even small systematic errors could become significant. Apart from the aspects discussed above, possible sources of errors include the finite accuracy of the GR waveform templates and the calibration of the GW detectors. We already have preliminary indications that combined posteriors from a large number of events could be dominated by such errors. Therefore, a careful characterization of various systematic errors is required before we accumulate a large number of detections to analyze. We leave this as future work.

\paragraph{Acknowledgments.} 
We are grateful to M.~Vallisneri and the anonymous referees for their insightful comments on the manuscript. We thank J.~Veitch, A.~Nagar and P.~Schmidt for assistance with the \textsc{LALInference}, IHES EOB and NR injection codes, respectively. We also thank the SXS Collaboration for creating a public archive of NR waveforms. We have benefited from useful discussions with K.~G.~Arun, A.~Buonanno, N.~Christensen, M.~K.~Haris, B.~R.~Iyer, S.~Kumar, A.~K.~Mehta, C.~Messenger, A.~Mukherjee, B.~S.~Sathyaprakash, M.~Vallisneri, C.~Van Den Broeck, S.~Vitale and several members of the LIGO Scientific and Virgo Collaboration's working group for strong gravity. Ar.~G., N.~K.~J.-M., and P.~A. acknowledge support from the AIRBUS Group Corporate Foundation through a chair in ``Mathematics of Complex Systems'' at ICTS. P.~A.'s research was, in addition, supported by a Ramanujan Fellowship from the Science and Engineering Research Board (SERB), India, the SERB FastTrack fellowship SR/FTP/PS-191/2012, and by the Max Planck Society and the Department of Science and Technology, India through a Max Planck Partner Group at ICTS. W.~D.~P. was partly supported by a Leverhulme Trust research project grant. C.~P.~L.~B. was supported by the Science and Technology Facilities Council. Computations were performed at the ICTS clusters Mowgli, Dogmatix, and Alice. This is LIGO document LIGO-P1700006.

\bigskip 

\appendix
\section{Calculation of the posterior of the parameters describing deviations from GR}
\label{app:posterior_derivation}
Here we describe the calculation of the posterior $P(\epsilon, \sigma \,|\, d)$ of the parameters $(\epsilon, \sigma)$ describing the fractional difference between the two independent estimates of the mass and spin of the remnant. We first make a change of variables in the joint posterior $\Pin(M^\textsc{i}_f, a^\textsc{i}_f\,|\, d)\Prd(M^\textsc{mr}_f, a^\textsc{mr}_f\,|\, d)$ from $\{M^\textsc{i}_f, a^\textsc{i}_f, M^\textsc{mr}_f, a^\textsc{mr}_f\}$ to $\{\epsilon, \sigma, \bar{M}_f, \bar{a}_f\}$ and then marginalize over $\bar{M}_f$ and $\bar{a}_f$.
Starting from the definitions given in equations~(\ref{eq:epsilon_sigma}) and~(\ref{eq:averages}), we find that
\begin{eqnarray}
M^\textsc{i}_f &= \left(1 + \frac{\epsilon}{2}\right)\bar{M}_f, \quad a^\textsc{i}_f &= \left(1 + \frac{\sigma}{2}\right)\bar{a}_f,\nonumber\\
M^\textsc{mr}_f  &= \left(1 - \frac{\epsilon}{2}\right)\bar{M}_f, \quad a^\textsc{mr}_f &= \left(1 - \frac{\sigma}{2}\right)\bar{a}_f,
\end{eqnarray}
so that the Jacobian of the transformation from $\{\epsilon, \sigma, \bar{M}_f, \bar{a}_f\}$ to $\{M^\textsc{i}_f, a^\textsc{i}_f, M^\textsc{mr}_f, a^\textsc{mr}_f\}$ is $\bar{M}_f\bar{a}_f$.
Thus, the final expression is
\begin{eqnarray}
P(\epsilon, \sigma \,|\, d) &= \int_0^1\int_0^{\infty} \Pin\left(\left[1 + \frac{\epsilon}{2}\right]\bar{M}_f, \left[1 + \frac{\sigma}{2}\right]\bar{a}_f\,\Big|\,d\right)\nonumber\\
&\quad\quad \times \Prd\left(\left[1 - \frac{\epsilon}{2}\right]\bar{M}_f, \left[1 - \frac{\sigma}{2}\right]\bar{a}_f\,\Big|\,d\right)~ \bar{M}_f\bar{a}_f ~ \dd\bar{M}_f\, \dd\bar{a}_f.
\end{eqnarray}

\bigskip 
\bigskip

\bibliographystyle{iopart-num}
\bibliography{TestGR_followup}

\end{document}